\documentclass{aa}  
\usepackage{graphicx}
\usepackage{afterpage}
\usepackage{txfonts}
\usepackage{xcolor}
\usepackage{soul}
\usepackage{subcaption}
\usepackage{caption}
\usepackage{natbib}
\bibpunct{(}{)}{;}{a}{}{,}
\usepackage[pdfencoding=auto,psdextra]{hyperref}
\hypersetup{
    colorlinks=true,
    linkcolor=blue,
    filecolor=magenta,      
    urlcolor=blue,
    citecolor=blue
}
\begin{document}

\title{Substellar initial mass function of Trumpler 14}

   \author{T. Rom
          \inst{1,2},
          E. Moraux\inst{2}, K. Mužić\inst{3}, M. Andersen\inst{4}, M. Schirmer\inst{5}, V. Almendros-Abad\inst{6}}

   \institute{University of Split, Faculty of Science, Ruđera Boškovića 33, 21000 Split, Croatia\\
              \email{trom@pmfst.hr}
         \and
             Univ. Grenoble Alpes, CNRS, IPAG, 38000 Grenoble, France 
        \and 
             Instituto de Astrofísica e Ciências do Espaço, Faculdade de Ciências, Universidade de Lisboa
        \and
             European Southern Observatory, Karl-Schwarzschild-Strasse 2, D-85748 Garching bei München, Germany
        \and
             Max-Planck-Institut f\"ur Astronomie, K\"onigstuhl 17, 69117 Heidelberg, Germany
        \and 
            Istituto Nazionale di Astrofisica (INAF) - Osservatorio Astronomico di Palermo, Piazza del Parlamento 1, 90134 Palermo, Italy }

   \date{Received 24 July 2025; accepted 25 November 2025 }

  \abstract
{Young, massive stellar clusters offer a prime setting to explore brown dwarf (BD) formation under high densities and intense UV radiation. Trumpler 14 (Tr 14), a $\sim$1 Myr-old cluster located at a distance of 2.4 kpc, and particularly rich in O-type stars, is an ideal target for such a study.}
% aims heading (mandatory)
{Our goal is to measure the initial mass function (IMF) in the young massive, high UV flux cluster.}
% methods heading (mandatory)
{We present the deepest study to date of the IMF in Tr 14 based on  Gemini Multi-Conjugate Adaptive Optics System/ Gemini-South Adaptive Optics Imager imaging. We construct the IMF using both the Besançon Galactic model and an observational control field from VISTA for background correction. Completeness was assessed using artificial-star tests and applied to the IMF derivation.}
% results heading (mandatory)
{We estimate the IMF down to the 20\% completeness limit found at $\sim 0.01\ M_\odot$. Using the control field-based IMF as our primary result, we find a slope of  $\alpha=0.14\pm0.19$ for masses between 0.01–0.2 $M_\odot$, and  $\alpha=1.72\pm0.04$ for 0.2–4.5 $M_\odot$, where $dN/dM \propto M^{-\alpha}$. The low-mass slope is largely influenced by the incompleteness-affected lowest bin; excluding it brings our results into agreement with those from other young clusters. The resulting median for the star-to-BD ratio in the $0.03-1\ M_\odot$ mass range is 4.0, with a 95\% confidence interval of 2.8-5.8. }
% conclusions heading (optional), leave it empty if necessary
{Our analysis reveals that Tr 14 hosts a relatively flat substellar IMF, but this is strongly influenced by the lowest-mass bin, which may be affected by incompleteness. When that bin is excluded, the IMF becomes consistent with those of other regions. The star-to-BD ratio falls within the usually observed $\sim$3–6 range, indicating that brown dwarfs with masses above $\sim$0.03 $M_\odot$ form with similar efficiency across environments. However, the relative lack of objects below this threshold suggests that the presence of an environment with both high stellar density and FUV flux may play a role in shaping the IMF by suppressing the formation of BDs at masses $<0.03\ M_\odot$. }


   \maketitle
%
%-------------------------------------------------------------------

\section{Introduction}

The study of star formation is of crucial importance in many aspects of astrophysics, for example, cosmological evolution at large scales or the formation of planetary systems at smaller scales. 
The prevailing view is that stars form within star clusters as a result of the gravitational collapse of molecular clouds \citep{kroupa2024initialmassfunctionstars, hennebelle2024physicaloriginstellarinitial}. Within these clusters, stars and brown dwarfs of varying masses emerge, spanning several orders of magnitude, making them an ideal environment for investigating the formation of low-mass stars and brown dwarfs (BDs).

The initial mass function (IMF) describes the distribution of stellar masses at birth and serves as a fundamental tool for comparing the outcome of star formation in various environments. Above 1 $M_\odot$, the IMF is often considered universal and follows the Salpeter power-law slope ($dN/dM \propto M^{-\alpha}$ with $\alpha=2.35$; \citealt{salpeter1995}), although a slightly shallower slope has been observed in some regions (see \citealt{bastian2010} for a comprehensive review). To describe the sub-solar range, \cite{kroupa2001} proposed a broken power-law with $\alpha=0.3$ in the $0.01-0.08\ M_\odot$ range and $\alpha=1.3$ in the $0.08-0.5\ M_\odot$ range. Additionally, \citep{chabrier2003,chabrier2005} provides an alternative description, suggesting a log-normal function ($dN/d\ \mathrm{log}m \propto e^{-\frac{(\mathrm{log}\ m - \mathrm{log}\ m_c)^2}{2\sigma^2}}$), 
with a peak at $0.25\ M_\odot$. 

Following the power-law description, observational studies of the low-mass IMF report a range of values for $\alpha$: $\alpha=0.67\pm0.23$ and $\alpha=0.29\pm0.22$ for Cha-I and Lupus 3, respectively for the mass ranges $0.02-0.2\ M_\odot$  and $0.04-0.2\ M_\odot$  {\citep{corona}}, $\alpha=0.7-1.1$ for NGC 2244 for $0.045-0.2\ M_\odot\ \mathrm{or}\ 0.045-0.4\ M_\odot$ \citep{victor2023}, $\alpha=0.51\pm0.15$ for RCW 38 for masses $0.02-0.2\ M_\odot$  {\citep{corona}}, $\alpha=0.18\pm0.19$ for masses $0.004-0.19\ M_\odot$ for $\sigma$ Orionis \citep{damian2023}, $\alpha=0.26\pm0.04$ for masses $0.011-0.4\ M_\odot$ for 25 Ori \citep{suarez2019}. These differences in values could arise from uncertainties affecting the IMF derivation, rather than intrinsic variations between clusters. Key factors influencing the determination of the IMF include assumptions about stellar age, cluster distance, theoretical models, membership selection methods, extinction law and varying mass ranges for the fits presented here \citep{scholz2013}.

The low-mass IMF is closely linked to BD production efficiency. The typical ratio of low-mass stars ($0.075-1\ M_\odot$) to BDs ($0.03-0.075\ M_\odot$) ranges from 2 to 6 \citep{muzic2015, muzic2017, andersen_2008, victor2023}. Several mechanisms have been proposed for the formation of BDs, including turbulent fragmentation \citep{padoan2002}, ejection from multiple systems \citep{reipurth2001}, formation in massive circumstellar disks \citep{stamatellos2009}, and photoevaporation of prestellar cores by OB-type stars \citep{whitworth2004}. However, studies of nearby regions suggest that most massive BDs likely form in a manner similar to stars (see \citealp{luhman_formation_2012} for a review), implying the necessity of examining more massive clusters to probe potential environmental variations of the IMF. 
Several of these theories predict that BD formation efficiency may depend on environmental factors such as stellar density, gas density, and the presence of massive stars \citep{luhman_formation_2012}. However, this remains an active area of research since it is not yet clear to which extent each of these factors may affect the low-mass IMF. 

One of the most massive star-forming regions in the Galaxy is the Carina Nebula molecular complex, containing young massive clusters: Trumpler 14 (Tr 14), Trumpler 15 and Trumpler 16 \citep{tapia2003}. Tr 14 contains $\sim 20$ OB-type stars \citep{shull2021gaia}, which could influence the IMF \citep{whitworth2004}. 
Moreover, we estimate the stellar density of Tr 14 to be $\sim 10^3$ stars/pc$^2$ (see Sect. \ref{sec:bd}), several times denser than the Orion Nebula cluster (ONC) as found in \cite{victor2023}.
Estimates of Tr 14’s age vary between <1 Myr and 5 Myr \citep{tapia2003, ascenso2007, hur2012, itrich2023}, with evidence for two distinct populations: a younger (<1 Myr) core and an older ($\sim$5 Myr) halo \citep{ascenso2007}. The latest spectroscopic analysis utilising MUSE IFU observations places the age of Tr 14 at $\sim$ 1 Myr \citep{itrich2023}, which we adopt in our study. The cluster distance is taken as $2.37\pm0.15$ kpc, based on Gaia EDR3 \citep{shull2021gaia}, in agreement with \citet{goppl_preibisch2022} from Gaia DR3 \citep{gaia_dr3, gaia_mission}. While \cite{ascenso2007} found a normal extinction law ($R_V=3.1$) with spatial variations, \cite{hur2012} suggested an anomalous extinction law towards Carina. The extinction law that we use throughout the paper is from \cite{extinction_law}, with assumed $R_V=3.1$. 

Previous studies \citep{ascenso2007, rochau, hur2012} of the IMF in Tr 14 have primarily focused on the stellar content, without robust coverage of the low-mass and substellar populations. For example, \citet{ascenso2007} detect sources down to K$_S\sim$18.5 mag but are only 90\% complete to K$_S\sim$14.5 mag, corresponding to $\sim 1-2\ M_\odot$ at 1 Myr depending on extinction and evolutionary models.  \citet{hur2012} report an observational limit of $\sim1.6\ M_\odot$. In contrast, \citet{rochau} reach significantly deeper, down to $\sim0.25\ M_\odot$. 
With its rich massive-star population and young age, Tr 14 provides a valuable setting to investigate BD formation and the effect of environmental factors in shaping the low-mass end of the IMF, therefore we did this work.

In this work, we present the most in-depth substellar survey of Tr 14 to date, using high-resolution near-infrared imaging from the Gemini-South Adaptive Optics Imager. Our analysis focuses on the central $\sim$$1.4'\times1.4'$ region of the cluster, corresponding to $\sim0.98\times0.98$ pc$^2$ for a distance of 2.4 kpc, aiming to characterize the low-mass IMF and evaluate the efficiency of BD formation in a high-radiation environment.
We present our observations, the reduction of data and the photometry in Sect. \ref{sec:obs}. Statistical determination of cluster membership is described in Sect. \ref{sec:clean}. The derivation of individual object mass and the initial mass function in Sect. \ref{sec:res}. Summary and conclusions are given in Sect. \ref{sec:fin}.
  
\begin{figure*}
    \centering
    \includegraphics[width=\textwidth]{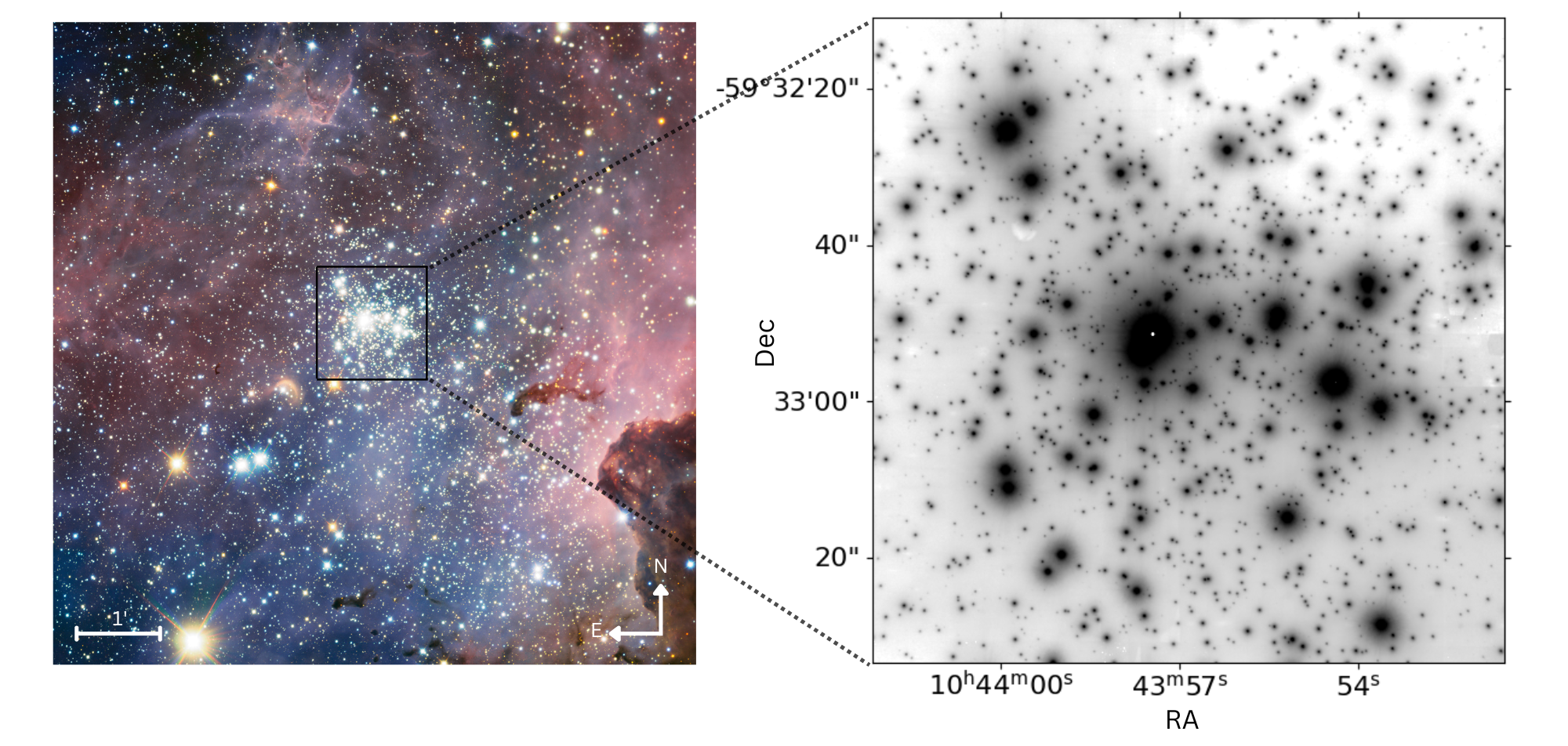}
    \caption{Left panel shows an image of Tr 14 in the Carina Nebula taken from \citet{carina}. The black square represents the  $\sim$$1.4'\times1.4'$ area studied in this work. The right panel shows the $J$-band image of Tr 14 taken with GSAOI. North is up and east to the left.}
    \label{fig:field}
\end{figure*}

%--------------------------------------------------------------------
\section{Observations, data reduction and photometry}
\label{sec:obs}

\subsection{Observations}
We observed the Tr 14 cluster in the $JHK_{\rm s}$ filters  using the Gemini-South Adaptive Optics Imager (GSAOI; \citealt{gsaoi1, gsaoi2}) assisted by the Gemini Multi-Conjugate Adaptive Optics System (GeMS; \citealt{gems1,gems2}), as a part of the program ID GS-2019A-DD-107 (PI: M. Andersen). Each exposure was 60s, and a total of 25, 15, and 15 exposures were obtained, in $J,H$ and K$_S$, respectively. Additionally, short exposures of 10s each were taken, with 10 exposures per filter. Three natural guide stars were utilised for the adaptive optics correction; their coordinates are given in Appendix \ref{appendix:ngs}. All observations were obtained under clear sky conditions. The image quality in the final images for $J,H$ and K$_S$ filters, is 0.14$''$, 0.13$''$ and 0.11$''$ respectively. 

\subsection{Data processing}
The images were processed using standard procedures with the {\tt THELI} pipeline \citep{schirmer2013}, and the GSAOI-specific adaptations in \cite{schirmer2015}. Pre-processing included dark subtraction, flat-fielding with domeflats, and the construction of background models using median-stacking of the dithered science images after source masking.

The astrometry and distortion correction were based on the Gaia-EDR3 reference catalogue. The final coadded image was resampled to the nominal GSAOI plate scale of $0.0197^{\prime\prime}$ per pixel. We show the $J$-band image in Fig~\ref{fig:field}. 

\label{sec:photo}
\subsection{PSF fitting photometry}\label{sub:sub1}
We use {\sc SourceExtractor} \citep{sextractor} and PSFEx \citep{psfex} to obtain the point-spread-function (PSF) model and then extract sources and measure the photometry using PSF-fitting. 
We test various parameters to identify the optimal PSF fitting, including BASIS$\_$NUMBER, PSF$\_$SIZE, PSFVAR$\_$KEYS, PSFVAR$\_$GROUPS, PSFVAR$\_$DEGREES, SAMPLE$\_$MINSN.  
PSFVAR$\_$KEYS and PSFVAR$\_$GROUPS represent the image parameters with which the PSF varies (e.g. X and Y pixel position in the image) and the polynomial group, respectively. PSFVAR$\_$DEGREES stands for the degree of the polynomial of the PSF. 
We use the same parameters consistently on all the images; the values are given in Table \ref{tab:parameters}.

Since our field is crowded, for each band, both in long and short-exposure images, we made two separate extractions, following the concept of "hot" and "cold" detection as outlined in \cite{sextractorhotcold}. With just one extraction we were not able to extract both the bright and the faint sources at the same time. Thus, the first extraction (called extraction 1 hereafter) parameters were optimized to gain more sources that are brighter and easier to resolve, and the second extraction (or extraction 2) was done to obtain fainter sources in crowded areas. The most significant parameters for extraction (BACK\_SIZE, BACK\_FILTERSIZE, and DETECT\_THRESH, DEBLEND\_MINCONT) affected which sources were detected in the images. The BACK\_SIZE parameter defines a cell size in which the local background is estimated using the mode value, while the BACK\_FILTERSIZE parameter smooths the background map by applying a median filter. The parameter DETECT\_THRESH dictates the threshold above the sky background in units of standard deviation of the background noise. The DEBLEND\_MINCONT parameter dictates the contrast for deblending multiple sources.\\

The obtained catalogues were cleaned for PSF\_MAG=99, FLAGS$>=$4, and SPREAD\_MODEL values deviating by more than 3$\sigma$ from the mean (see Table~\ref{tab:parameters} for the exact values). PSF\_MAG=99 represents a bad PSF magnitude extraction, e.g. a saturated source, while SPREAD\_MODEL excludes all the deviated and ellipsoidal sources. FLAGS>=4 rejects sources on the edges of the images and with saturation, or corrupted photometric aperture (see {\sc Source-extractor} documentation for more details; \citealp{sextractor}).

\subsection{Photometric calibration}
\begin{figure}[h]
    \centering
    \includegraphics[width=0.47\textwidth]{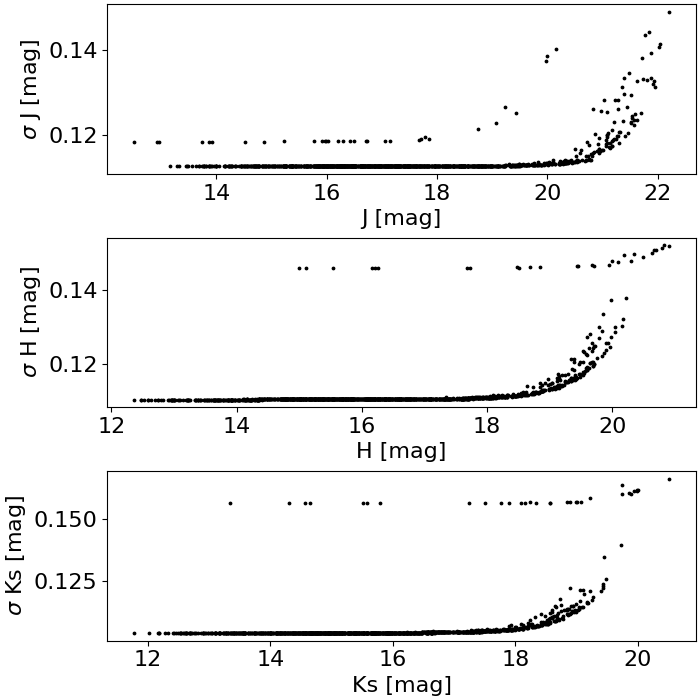}
    \caption{Photometric uncertainties of our catalogue as a function of magnitude. In all bands, two prominent sequences are visible, originating from the long- and short- exposure observations.}
    \label{fig:photcal}
\end{figure}

Once we obtained the photometry (Sect. \ref{sub:sub1}), the extraction 1 and 2 catalogues were combined and then calibrated using the catalogue from \citet{ascenso2007}. The calibration catalogue contains data from two different instruments, with different spatial resolutions. We used only the low-spatial resolution instrument, to avoid possible errors (e.g. matching 2 stars from our catalogue to 1 star in the calibration catalogue) in the cross-matching process, and matched it with our catalogue with a tolerance of 0.3$''$.  Moreover, for calibration purposes, we excluded the sources in our catalogue that have another source within 0.5$''$, to avoid mismatches between the two catalogues. The calibration catalogue from \citet{ascenso2007} has been calibrated against the Two Micron All Sky Survey (2MASS)  photometric system \citep{2mass}. We calibrated the photometry using the zero-points only and neglected the colour-terms due to them being consistent with zero within the errors. To derive the zero-point (ZP) values for photometric calibration, we take the difference of the instrumental magnitudes from our data and reference magnitudes from the calibrated catalogue, taking measurement errors into account and using 3-sigma clipping. 
The final ZP was calculated as a weighted average of the surviving data points, with weights determined by the inverse squared errors of both datasets, providing the ZP and its associated uncertainty. As a result, for the long exposure, we get ZP$_J=25.97\pm0.11$ mag, ZP$_H=26.18\pm0.15$ mag, ZP$_{\mathrm{K}_S}=25.64\pm0.16$ mag, and for the short exposure ZP$_J=25.99\pm0.11$ mag, ZP$_H=26.33\pm0.11$ mag, ZP$_{\mathrm{K}_S}=25.75\pm0.10$ mag.
The photometric uncertainties in individual $J$, $H$, and $K_S$ catalogues were calculated by combining the uncertainties of the zero-points and the measurement uncertainties supplied by {\sc Source-Extractor}, and are shown in Fig.~\ref{fig:photcal}.

The final catalogue for each band was obtained by matching the long and short-exposure catalogues. For the $J$-band, we used 3-pixel separation, while the rest had 2.5-pixel separation due to the smaller full-width-at-half-maximum of the sources. For the sources that appear in both catalogues, we kept the values with the lower magnitude error. Finally, we cross-matched the catalogues in the three photometric bands using 2.5-pixel separation. We cut off 400 pixels off each edge to exclude sources with lower signal-to-noise ratio. The final catalogue contains 1207 sources with photometry in all three photometric bands, 3 additional sources in only $J$ and $H$ bands, 12 sources in only $J$ and K$_S$ bands, and 356 sources in only $H$ and K$_S$ bands. The first three lines of the catalogue are given in the Table \ref{tab:cat} and the full catalogue is provided electronically.

\begin{table*}
\captionsetup{justification=raggedright,singlelinecheck=false}
\caption{JHKs catalogue of Tr 14.}
     \label{tab:cat}
\centering
    \begin{tabular}{cccccccc}
    \hline \hline
         RA [h:m:s] & Dec [d:m:s]  & $J$ [mag] & $e_J$ [mag] & $H$ [mag] & $e_H$ [mag] & $K_s$ [mag] & $e_{K_s}$ [mag] \\
         \hline
         10:43:54.93 & –59:33:31.28 & 14.15 & 0.11& 13.19 & 0.11 & 12.90 & 0.10 \\
         %\hline
         10:43:54.66 & –59:33:31.97 & 15.83 & 0.11 & 14.76 & 0.11 & 14.24 & 0.10 \\
         %\hline
         10:43:55.41 & –59:33:30.78 & 16.09 & 0.11 & 14.99 & 0.11 & 14.39 & 0.10 \\
         \hline
    \end{tabular}
    \tablefoot{The full table is available in the electronic form on CDS.}  
\end{table*}

\subsection{Completeness}
\begin{figure}[h]
    \centering
    \includegraphics[width=0.47\textwidth]{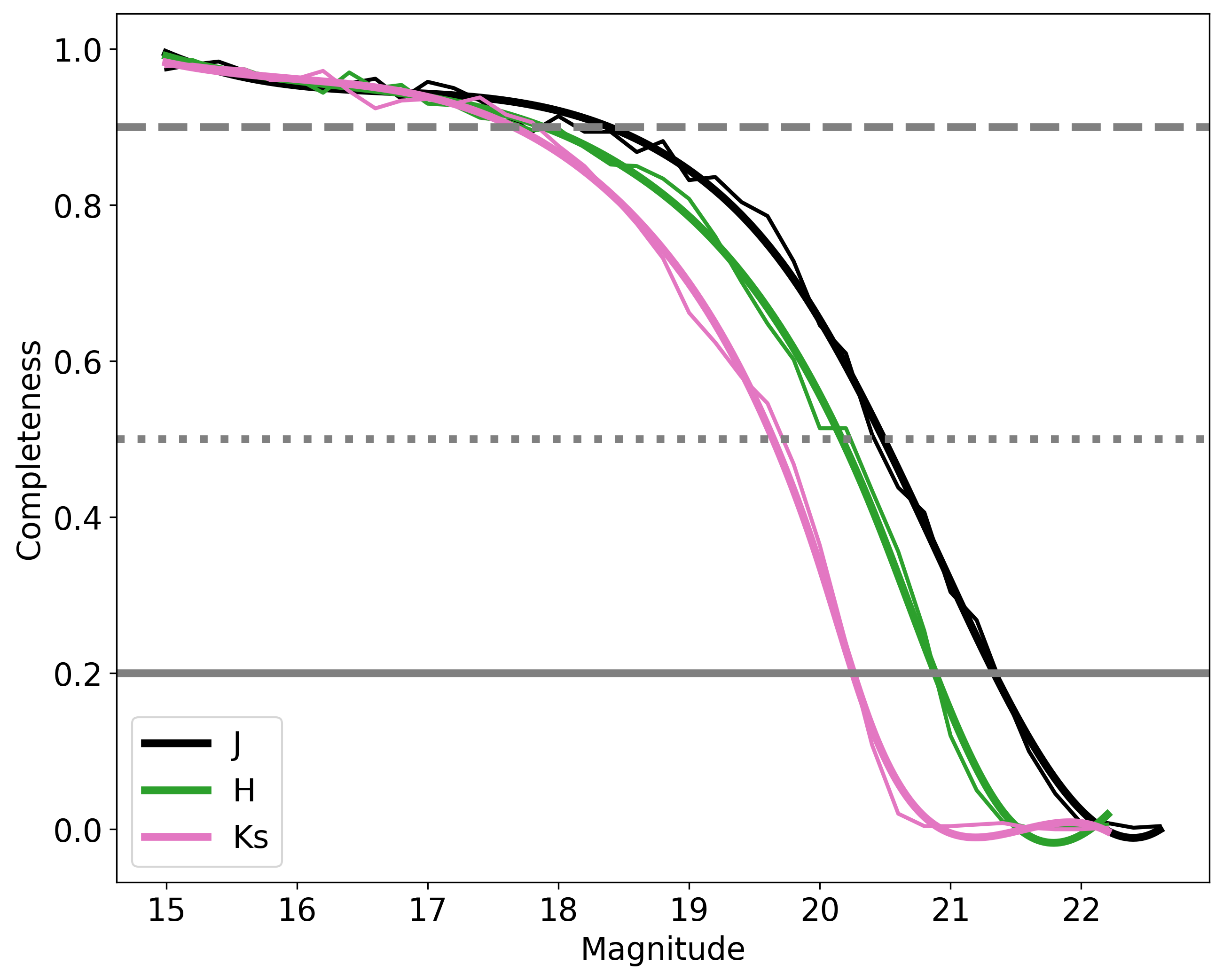}
    \caption{Completeness of the photometry, obtained using the artificial star test. Different  {coloured} lines represent corresponding photometric bands. The dashed grey line represents the 90\% completeness limit, while the dotted grey line represents the 50\% completeness limit and the solid grey line the 20\% completeness limit. The thin lines show the completeness curves calculated at the 0.2 mag step, while the thick lines were smoothed using \textit{ UnvariateSpline} of scipy.interpolate sub-package.}
    \label{fig:completeness}
\end{figure}

We evaluated the photometric detection completeness through artificial star experiments. Artificial stars were randomly positioned in our images using the \textit{source} function of Artificial Stellar Populations ({\sc ArtPop}) Python package \citep{artpop}. We injected the stars using the PSF model obtained with PSFEx and the parameters listed in Sect. \ref{sub:sub1}. We inserted 50 stars at a time and repeated it 10 times, to improve statistics, in steps of 0.2 mag. The extraction used the same parameter values, PSF model, and procedure as in the photometric calibration. Fig. \ref{fig:completeness} shows the fraction of recovered stars in comparison to the inserted artificial ones, as a function of magnitude. 
The 20\% completeness limit is at $J=21.3$\,mag, $H=20.9$\,mag and K$_S=20.3$\,mag. This corresponds to the mass of $\sim$0.01 $M_\odot$, assuming the distance of 2.37 kpc, 1 Myr for the age and combined PARSEC and BT-Settl model (see Sect. \ref{sec:mass}), and with the derived median extinction of the cluster, $A_V=2.2$\,mag (see Sect. \ref{sec:mass}). 
The 90\% completeness limit is at $J=18.4$ mag, $H=17.9$ mag and K$_S=17.7$ mag, which corresponds to masses of 0.06 - 0.08 $M_\odot$. Table \ref{tab:clim} summarizes our results for the 20\%, 50\% and 90\% limits.  Previously deepest data in Tr 14 are down to $\sim 0.065 M_\odot$ from \cite{itrich2023}, at 30\% completeness and not including the core of the cluster.

\begin{table}[]
 \caption{Magnitudes and masses that correspond to specific completeness limits using the 1 Myr combined PARSEC and BT-Settl (\citealt{bressan2012parsec, chen2014improving, baraffe2015new}) isochrone at 2.37 kpc and $A_V=2.2$ mag.}
    \centering
    \begin{tabular}{cccc}
    \hline\hline
         & 20\% & 50\% & 90\% \\
        \hline
        $J$ [mag] & 21.3 & 20.5 & 18.4 \\
        $H$ [mag] & 20.9 & 20.2 & 17.9  \\
        $K_S$ [mag] & 20.3 & 19.6 & 17.7  \\
        $M/M_{\odot}$ & 0.01 & 0.02 & 0.06-0.08\\
        \hline
    \end{tabular}
    \label{tab:clim}
\end{table}

\section{Cluster membership}
\label{sec:clean}
\begin{figure*}[h!]
    \centering
    \includegraphics[width=0.9\textwidth]{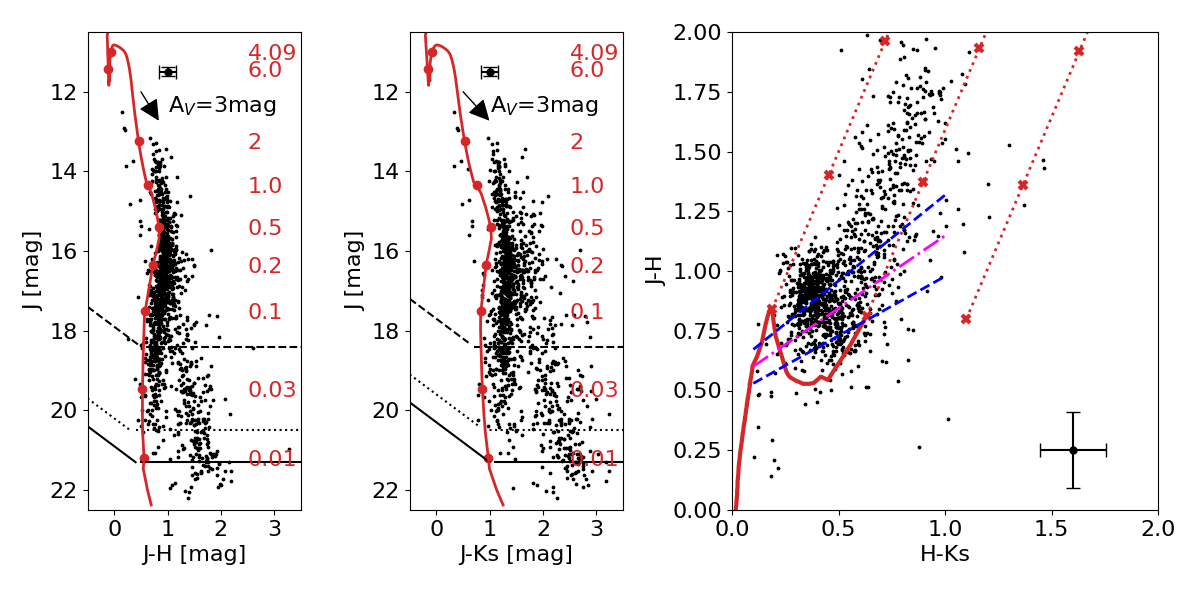}
    \caption{Left and central panel show the CMDs. The mean uncertainty is in the upper right part of the panels. The red solid line represents the 1 Myr isochrone combined from the PARSEC and BT-Settl models at 1 $M_\odot$ (\citealt{bressan2012parsec, chen2014improving, baraffe2015new}). The reddening vector of Av=3 mag is shown  {\citep{extinction_law}}. The dashed black line in the first two panels represents the 90\% completeness limit, while the dotted black line represents the 50\% completeness limit and the solid black line the 20\% completeness limit. The right panel represents the  colour- colour diagram (CCD). The solid red line is the isochrone as in the CMDs. The dashed red lines are parallel to extinction vectors, while the red crosses represent Av=0, 5 and 10 mag. The dashed-dotted magenta line represents the T-Tauri stars locus with the corresponding uncertainties as blue dashed lines \citep{meyer+1997}, transformed to 2MASS system using \cite{carpenter_color_2001}. The average uncertainty is in the lower-right corner in the right-most panel, in other panels it is up right. The starting points for the left and middle reddening vector are determined by the masses in the model (<0.5 $M_\odot$ and >=0.005 $M_\odot$, respectively) while the right reddening vector is from \cite{muzic2017}, adapting the locus of Herbig AeBe stars from \cite{Hernandez_2005}.}
    \label{fig:CMD_CCD}
\end{figure*}

To examine the cluster population, we show two  colour-magnitude diagrams (CMDs) of the observations in the first two panels of Fig. \ref{fig:CMD_CCD}. The cluster sequence can be fairly well distinguished, with  colours $J-$K$_S\approx1-1.7$ mag, but there is a sparsely populated foreground population (to the left of the cluster sequence), and background contamination (to the right) not following the shape of the isochrone and located at the fainter end of the CMD, both of which probably partially overlap with the cluster population.  The background population seen in the CMDs in the cluster direction appears to be significantly redder than the cluster sequence. The cluster seems to be located in front of a molecular cloud as seen from the Earth \citep{ascenso2007} and it is therefore expected that the background sources suffer larger extinction.

 Therefore, in our field-of-view, we see both the cluster members and field stars contaminating our data set. We have to be able to differentiate between the two to do a proper analysis of the cluster. Thus, we need to perform a membership determination. In this section, we perform the decontamination of the observations using two independent control field (CF) datasets: the Besançon Galaxy model\footnote{\url{https://model.obs-besancon.fr/}}, and a CF obtained using VIRCAM data from the Visible and Infrared Survey Telescope for Astronomy (VISTA) telescope \citep{vista}. 
 The setups for the CFs are explained in Sect. \ref{sec:de-model} and \ref{sec:vista}, respectively, followed by the common decontamination procedure in Sect. \ref{sec:stat}.

\subsection{Decontamination setup with a Galaxy model}
\label{sec:de-model}

To address the contamination with the Besançon Galaxy model presented in \cite{bgm},  as an input we use the central position ($l=287.4107^\circ,\ b=-00.5794^\circ$) of our cluster  {with} an area 24.5 times larger than our cluster area on the sky, which is scaled later in the process of decontamination (Sect. \ref{sec:stat}) with the photometric errors given as in the Appendix \ref{appendix:besancon}. 
We take this size of the area to get a significant number of the expected field stars. The output of the obtained  Besançon Galaxy model contains the $J$-band magnitude, $J-H, J-K$, and $H-K$  colours, age, mass, coordinates in right ascension and declination, distance and extinction. We transform the model magnitudes from the Koornneef system to 2MASS using the relations given in \citet{carpenter_color_2001}.  

 Since the Besançon Galaxy model provides the distances of the simulated stars, we divide the model output into foreground and background populations by comparing those values with the cluster’s location at 2.37 kpc.The foreground population will be used for decontamination as is, i.e. without applying any extinction. For the background population, we estimate the values of extinction by reddening the CF photometry until it coincides with the background contaminants seen in the cluster CMD (see Fig. \ref{fig:CF}). We thus decontaminate the cluster from the background after reddening it with the extinction values of $A_V=5,6,$ and 7 mag. The decontamination procedure, following the setup described here, is detailed in Sect. \ref{sec:stat}. 

\subsection{Decontamination setup with VISTA}
\label{sec:vista}

With the 4-m VISTA of ESO, \cite{vista} mapped an area of 6.7 deg$^2$ around the Carina Nebula in the near-infrared $J$-, $H$-, and K$_S$-bands,  {a portion of which we use as a CF}.
To address the contamination with the CF, we take the shape of a rectangle centred at 10h40m48.00s RA, -59°31'12.00" Dec, and 24.5 times bigger than the size of the cluster area. The CF is $\sim$  0.4° away from our cluster and they have a similar galactic latitude ($b_{\rm CF}=-1.118^{\circ},\ b_{\rm Tr 14}=-0.575^{\circ}$). We take this position and size to get a significant number of field stars. In order to separate the foreground and background populations of the CF we use Gaia parallaxes. 

We cross-match our catalogue to Gaia DR3 (\citealt{gaia_mission, gaia_dr3, gaia_cat}) to obtain the proper motions and parallaxes and we perform two cuts.  Firstly, we cut in proper motion in right ascension and declination, keeping only the most probable members of the cluster which should behave similarly in the proper motion diagram ($PM_{\mathrm{R.A.}}=-6.49\pm0.60$ mas yr$^{-1}$, $PM_{\mathrm{Dec}}=2.14\pm0.58$ mas yr$^{-1}$).  We do this by applying sigma clipping using a threshold of 3$\sigma$ with a maximum of 5 iterations, to iteratively exclude outliers that deviate significantly from the central distribution. This ensures that the derived statistical parameters reflect the intrinsic kinematic properties of the population, minimizing the impact of contaminants or measurement errors.  Secondly, we  {apply an additional brightness cut on the subset obtained from the first selection}, keeping only the brightest (<16 mag) objects to calculate the weighted mean parallax and parallax error.  This value is then used to identify and reject foreground objects in the Tr 14 catalogue based on a parallax zeta score > 2. We calculate the zeta score with the expression $\zeta=\mid \omega - \Bar{\omega} \mid / \sqrt{\sigma^2_\omega+\sigma^2} $, where $\omega$ and $\sigma_\omega$ are the parallax and the uncertainty for each star, while the $\Bar{\omega}$ and $\sigma$ are the weighted mean and uncertainty of the whole sample.   Finally, we have a dataset of our cluster that contains probable cluster members as well as the background contaminants.

\begin{figure*}
    \centering
    \includegraphics[width=\linewidth]{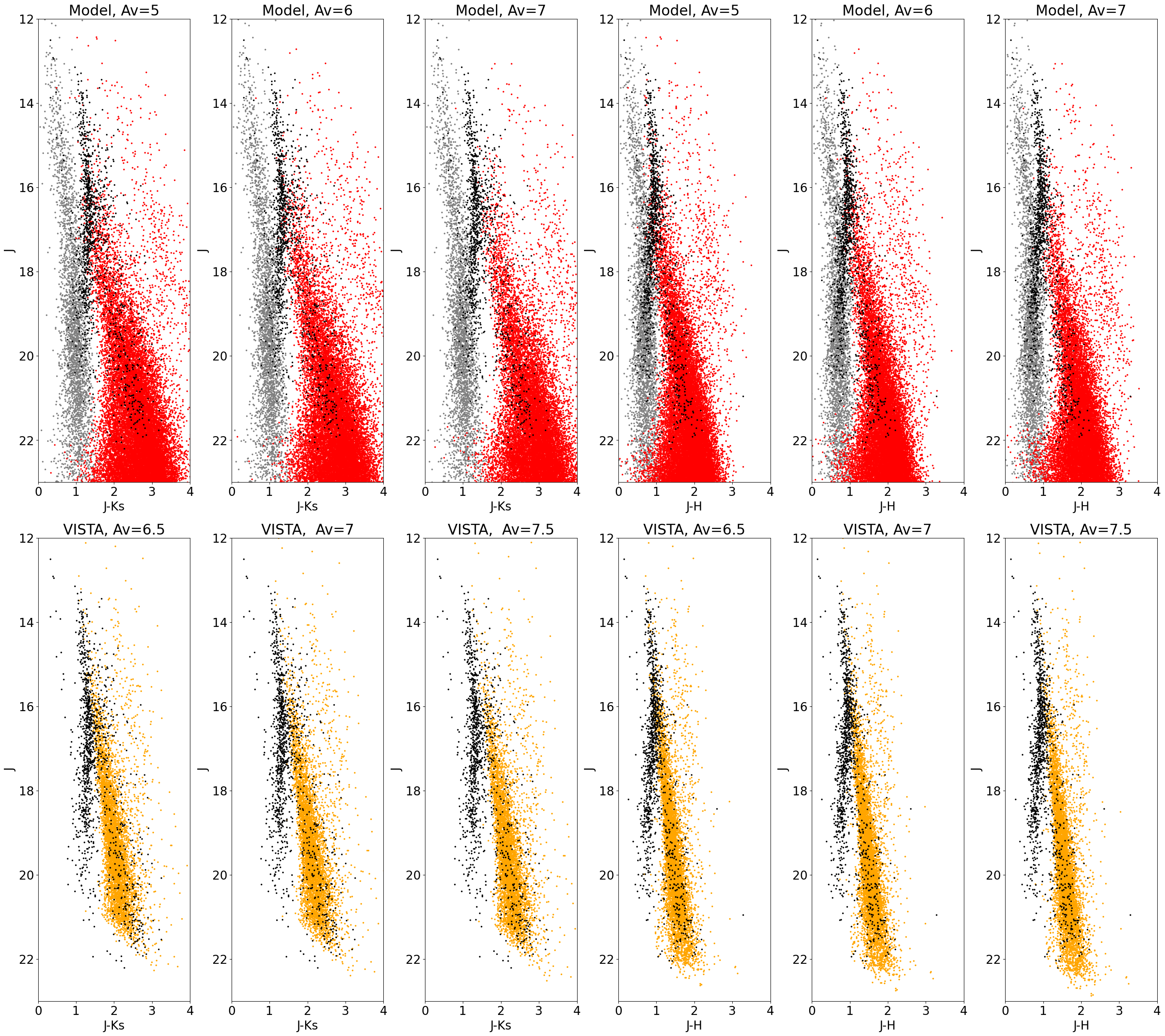}
    \caption{CMDs for the Besançon model in the upper panels and VISTA CF in the lower panels. The black dots show all the objects in the line of sight of Tr 14 in all panels. The grey dots are the model foreground, while the red ones are the reddened model background. The amount of reddening is given above each corresponding panel. The orange dots represent the reddened CF. Again, the amount of reddening is given above each panel. Each extinction is chosen to decontaminate the most amount of background sources.}
    \label{fig:CF}
\end{figure*}

\begin{figure*}[h]
    \centering
    \includegraphics[width=\textwidth]{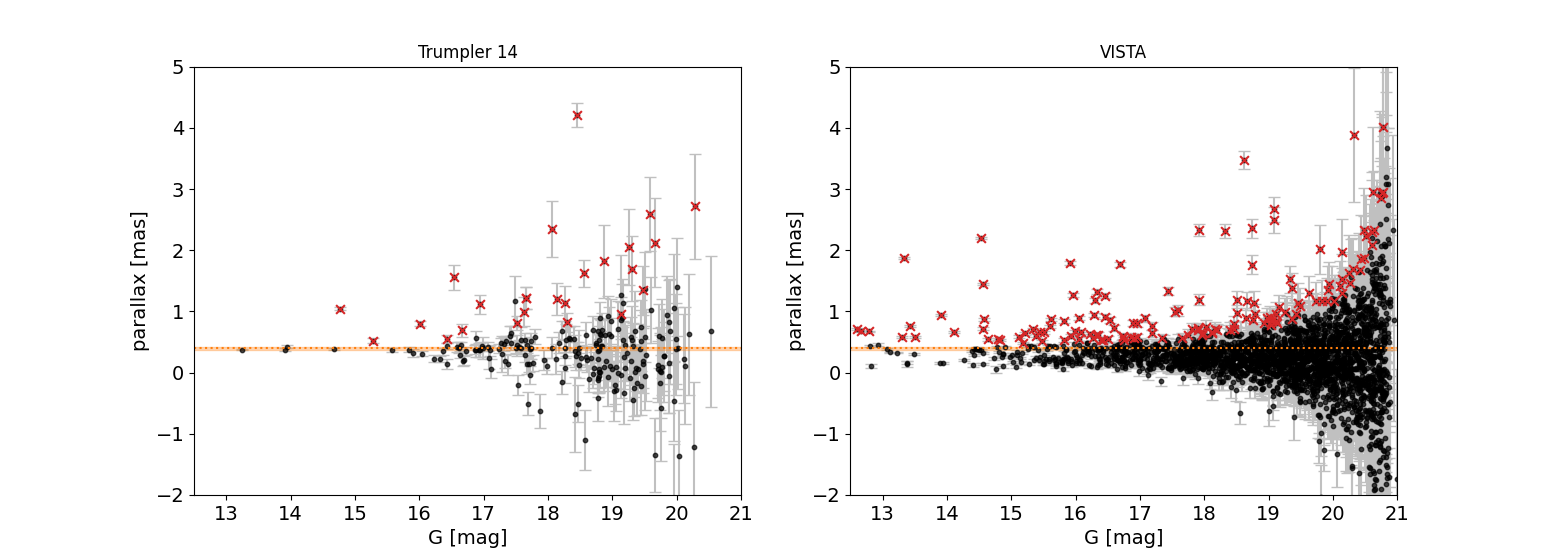}
    \caption{Figure shows the parallax versus the G magnitude band of Tr 14 (left panel) and VISTA CF (right panel). In both panels red crosses represent the objects that are rejected as a part of the foreground of Tr 14 (left panel) and CF (right panel). Black dots represent the objects of the cluster (in the left panel) and CF (in the right panel), while the  grey errorbars are parallax errors. To remove the foreground we set the zeta score for the parallax >2. The orange dashed line represents the mean cluster parallax and the orange rectangle is $\pm\sigma$ of the mean cluster parallax, in both panels.}
    \label{fig:parallax}
\end{figure*}

We then make the same cut  based on the parallax zeta score in the VISTA CF catalogue, that we previously cross-matched with Gaia DR3, to keep only the subset that corresponds to a probable background.  We show the results of these cut procedures in Fig. \ref{fig:parallax}.  To this VISTA CF catalogue with removed foreground objects, we apply different amounts of extinction ($A_V=6.5,7,$ and 7.5 mag, see Fig. \ref{fig:CF}) which we then use in the procedure explained in Sect. \ref{sec:stat}. 

\subsection{Statistical determination of  {membership}}
\label{sec:stat}
To statistically determine the cluster membership,  {after the independent setups for the CFs in Sect. \ref{sec:de-model} and Sect. \ref{sec:vista},} we  {apply} the procedure of \citet{muzic+2019}, which is summarized as follows:

\begin{enumerate}
    \item The CMD  containing the cluster and the CF, as in Fig. \ref{fig:CF} (upper panels if we use the Besançon model as the CF and lower panels if we use VISTA as the CF), is divided in a grid with cell sizes $\Delta \rm{ {colour}}$ and $\Delta \rm{magnitude}$.
    \item For each cell, the expected number densities of stars in the cluster field-of-view and in the CF (either from the Besançon model or from VISTA) are calculated. The densities are then scaled for different on-sky areas of the cluster and the CF.
    \item The expected number of field objects in each cell is then randomly removed from the cluster population.
\end{enumerate}

\begin{figure}
    \centering
    \begin{subfigure}{0.47\textwidth}
        \includegraphics[width=\textwidth]{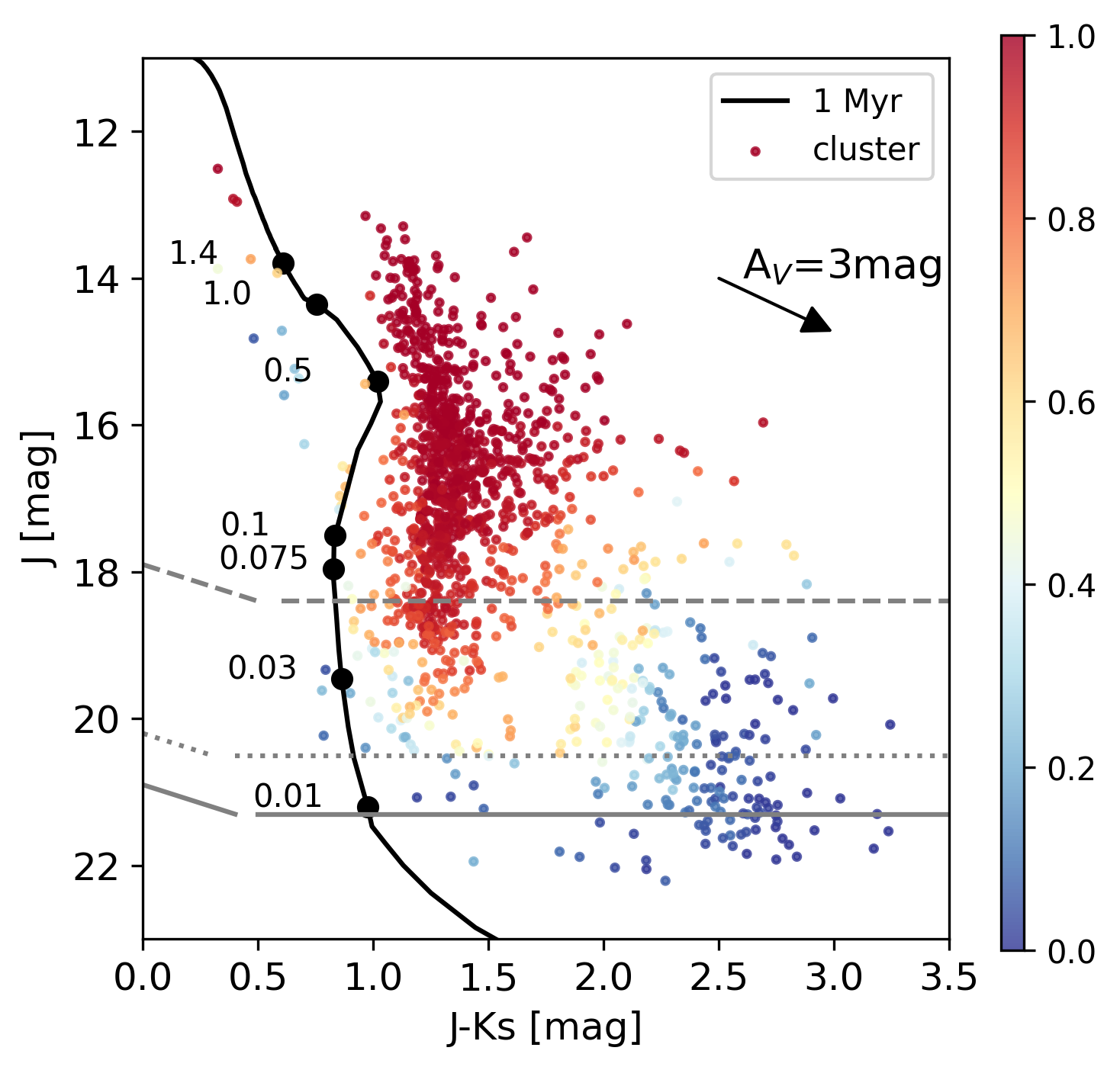}
        \label{fig:memb_model}
    \end{subfigure}
\hspace{-1cm}
    \begin{subfigure}{0.47\textwidth}
        \includegraphics[width=\textwidth]{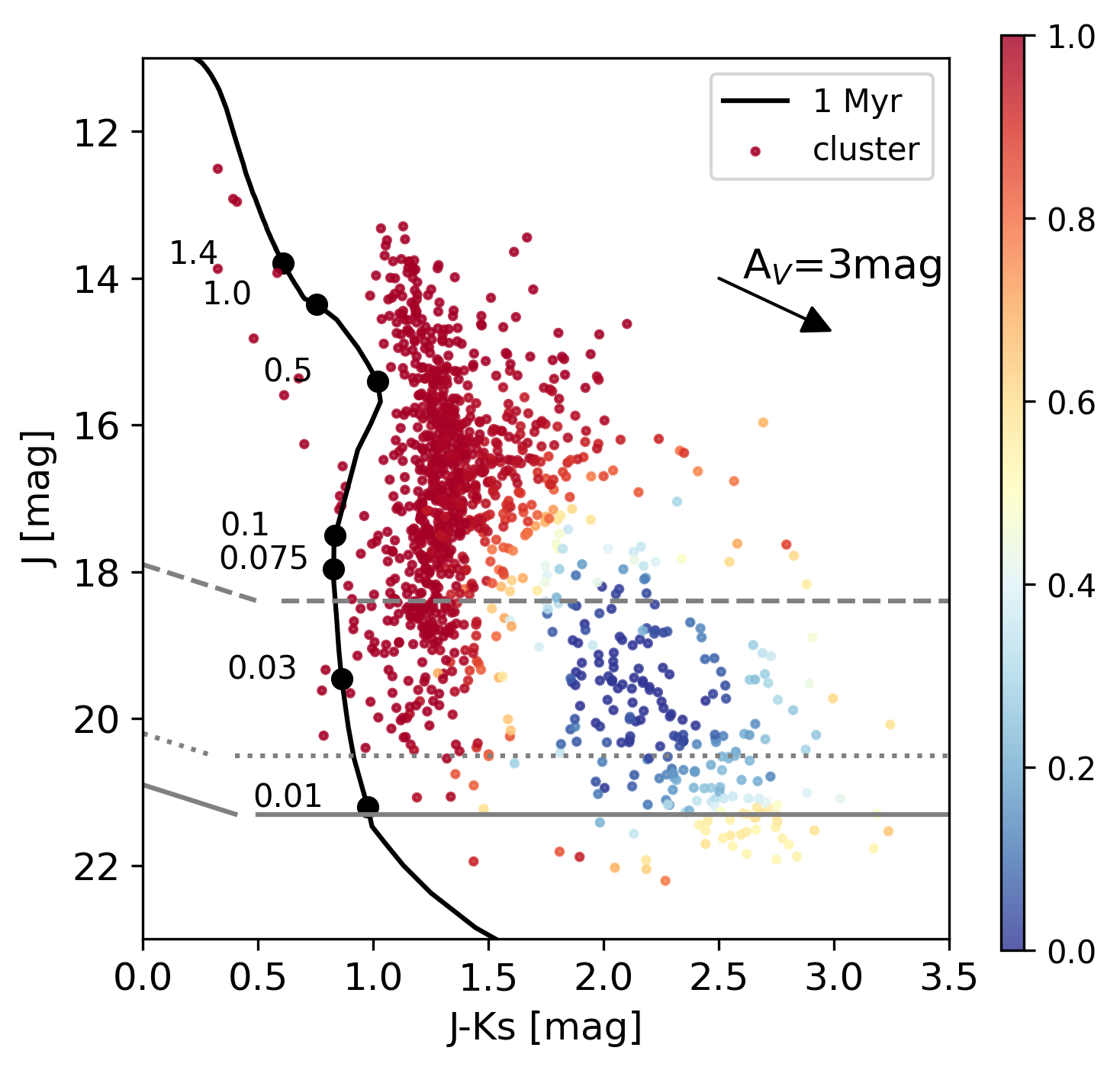}
        \label{fig:memb_vista}
    \end{subfigure}
     \caption{Cluster membership probability using the model (upper panel) and VISTA (lower panel) as a CF in the process of decontamination. The solid black line represents the 1 Myr isochrone combined from the PARSEC and BT-Settl models at 1 M$_\odot$ \citep{bressan2012parsec, chen2014improving, baraffe2015new}. The reddening vector of A$_V$=3 mag is shown. The dashed grey line represents the 90\% completeness limit, while the dotted grey line represents the 50\% completeness limit and the solid grey line the 20\% completeness limit.} 
    \label{fig:membership}
\end{figure}

Each object in the CMD is represented with a Gaussian probability distribution in the space of  {colour} and magnitude, with the width being determined by the photometric uncertainties. The part of the Gaussian that lies within the cell is then integrated and summed to get the expected number of objects for each cell. Cell sizes are determined in a way to have good statistics of cluster members and CF objects in each cell. We then remove from the cluster field as many sources as computed in the control field in each cell. The specific objects to be removed are chosen randomly. We used different cell sizes with $\Delta \rm{magnitude}=(0.4,0.5,0.6)$ mag, which results in $N_{mag}=3$ different magnitude bin sizes and $\Delta \rm{ {colour}}=(0.3,0.4)$ mag, yielding $N_{col}=2$  colour bin sizes. Additionally, to improve accuracy for the objects laying at the edges of our cells, for each magnitude-colour combination, we shift the cell center for 0.0, 1/3, -1/3 of the cell size in each dimension. This creates $N_{s}=3\cdot3=9$ cell center configurations.  We repeat the decontamination procedure for each combination of various parameters, namely cell sizes and cell centers, as well as the applied reddening. We look at $N_{cmd}=2$ CMD types ($J,J-K,J-H)$ for $N_{A_V}=3$ extinction values as shown in Fig. \ref{fig:CF}. This results in $N_{mag}\cdot N_{col}\cdot N_{s} \cdot N_{cmd} \cdot N_{A_V}=N_L=$ 324 lists of potential cluster members which contain IDs of catalogue sources that we later employ in the procedure for the initial mass function determination of the cluster. We calculate the probability for each star in the catalogue to be a cluster member by counting the number of times it appears in all lists and then dividing by the total number of lists.  {This methodology follows the similar approach previously applied in other young clusters such as NGC 2244 (e.g. \citealt{muzic+2019}), and has been validated through spectroscopic analysis by \citet{victor2023}, demonstrating its robustness in separating cluster and field populations.} The resulting probabilities are shown in Fig. \ref{fig:membership} where we use the VISTA (bottom panel) and the Besançon model (upper panel) as CFs, respectively, in the decontamination method. In the bottom panel, the foreground population appears as high-probability members since for the decontamination with VISTA we cross-match our catalogue with Gaia, which is much shallower than our photometry, thus the sample is limited.  {We then remove the foreground objects, which are known to contribute to contamination based on their Gaia parallaxes, as explained in Sect. \ref{sec:vista}}

\section{Results and discussion}
\label{sec:res}
\subsection{Extinction and masses}
\label{sec:mass}
To derive the IMF of our cluster, we have to estimate the masses of the cluster members. We use evolutionary models to get these values.
To obtain extinction and masses, we de-redden our sources to the $J$, $J-H$ isochrone in the CMD, as in \cite{muzic2017}. The full CMD and CCD are shown in Fig. \ref{fig:CMD_CCD}. We combined the PARSEC and BT-Settl models at 1 $M_\odot$ to a 1 Myr isochrone (\citealt{bressan2012parsec, chen2014improving, baraffe2015new}) as in \cite{muzic+2019}. 

First, we generate 1000 magnitude values for each source assuming a normal distribution, where the standard deviation is the photometric uncertainty. We check if the source, with the uncertainties is inside, or crossing the model limits (red solid and red dotted lines in Fig. \ref{fig:CMD_CCD} in the CCD), else the mass is not derived. The extinction is then derived by de-reddening in the CMD to the 1 Myr isochrone in these 1000 runs. The mass is obtained by interpolating the mass-magnitude relation corresponding to the model isochrone. 

\begin{figure*}
    \centering
    \includegraphics[width=\textwidth]{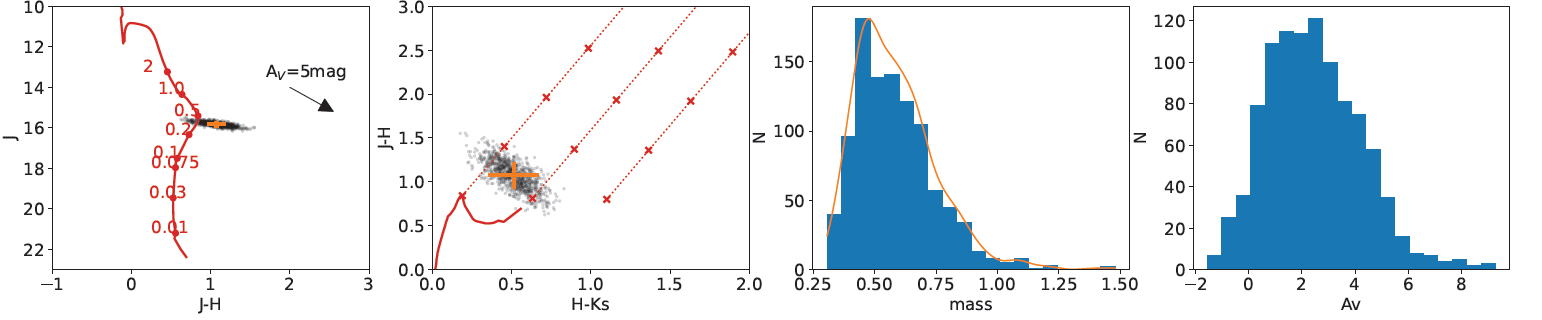}
    \caption{Example of mass derivation for one of the sources. The first two panels display the 1 Myr isochrone (red solid line) combined from the PARSEC and BT-Settl models at 1 $M_\odot$ mass (\citealt{bressan2012parsec, chen2014improving, baraffe2015new}) and the source with its error bars (orange solid cross). The grey dots represent the generated magnitude points from 1000 runs. In the first panel a line parallel to the extinction vector would show where the mass lies on the model. In the second panel, the red dotted lines are parallel to the reddening vectors. Red crosses mark the extinction in 5 mag steps from 0 to 15 mag. The third and the fourth panel display the mass and extinction distribution, respectively, obtained from 1000 runs.}
    \label{fig:mass_ext}
\end{figure*}

An example of the mass and extinction derivation procedure of one of the sources is shown in Fig. \ref{fig:mass_ext}. The right extinction line represents the limit, where on the left, we would expect to see T-Tauri \citep{meyer+1997} and Herbig AeBe stars \citep{Hernandez_2005}.  However, as argued in \citet{muzic+2019}, the excess coming from the circumstellar disc of the classical T-Tauri or Herbig AeBe stars appears to have a negligible impact on the IMF, thus we do not attempt to correct for it. The solid red line in Fig. \ref{fig:mass_ext} represents the isochrone, while the orange cross represents the source with its uncertainties. Additionally, mass and extinction histograms are plotted, with the orange line representing the Gaussian kernel-density estimated mass distribution. In these runs, it is possible to obtain negative $A_V$, however, this is not physical. To correct for it, without skewing the distributions, only data points with $A_V>=0$ are used to estimate the IMF.

As a result of this procedure, we get the mass distributions, median mass, median and mean extinction with its standard deviation for each source in the catalogue. The mass distributions are later used to obtain the IMF.

\label{sec:imf}
\subsection{Initial mass function}
We aim to derive the IMF from a set of mass distributions, obtained as described in Sect. \ref{sec:mass}, using the lists containing IDs of catalogue sources that remain after the decontamination procedure described in Sect.~\ref{sec:clean}. 

Each star's mass is represented by a mass distribution which is smoothed using a Gaussian KDE. From the mass distribution of each star, we draw a sample of $N_1=100$ values per object, resulting in 100 different mass distributions for the entire sample. Next, for each of the $N_1$ resampled sets, bootstrap (random samplings with replacement) is performed $N_2=100$ times to account for statistical fluctuations. This results in $N_1\cdot N_2$ individual mass distributions. During each iteration, we evaluate the IMF as a histogram with variable bin sizes, chosen to contain roughly the same number of objects and save it. The same bins are used in each iteration. This mass function is corrected for completeness by dividing each star in the bin individually with its corresponding completeness value. We exclude all the sources with completeness $<=20\%$ from the calculation of the IMFs because we do not want to be biased by the faint objects that have low completeness. To ensure the robustness of the derived IMF, we then average the $N_1\cdot N_2 \cdot N_L$ IMFs obtained and calculate the standard deviation to estimate the uncertainty of each bin. 

\begin{figure*}
\centering
    \begin{subfigure}{0.47\textwidth}
        \includegraphics[width=\textwidth]{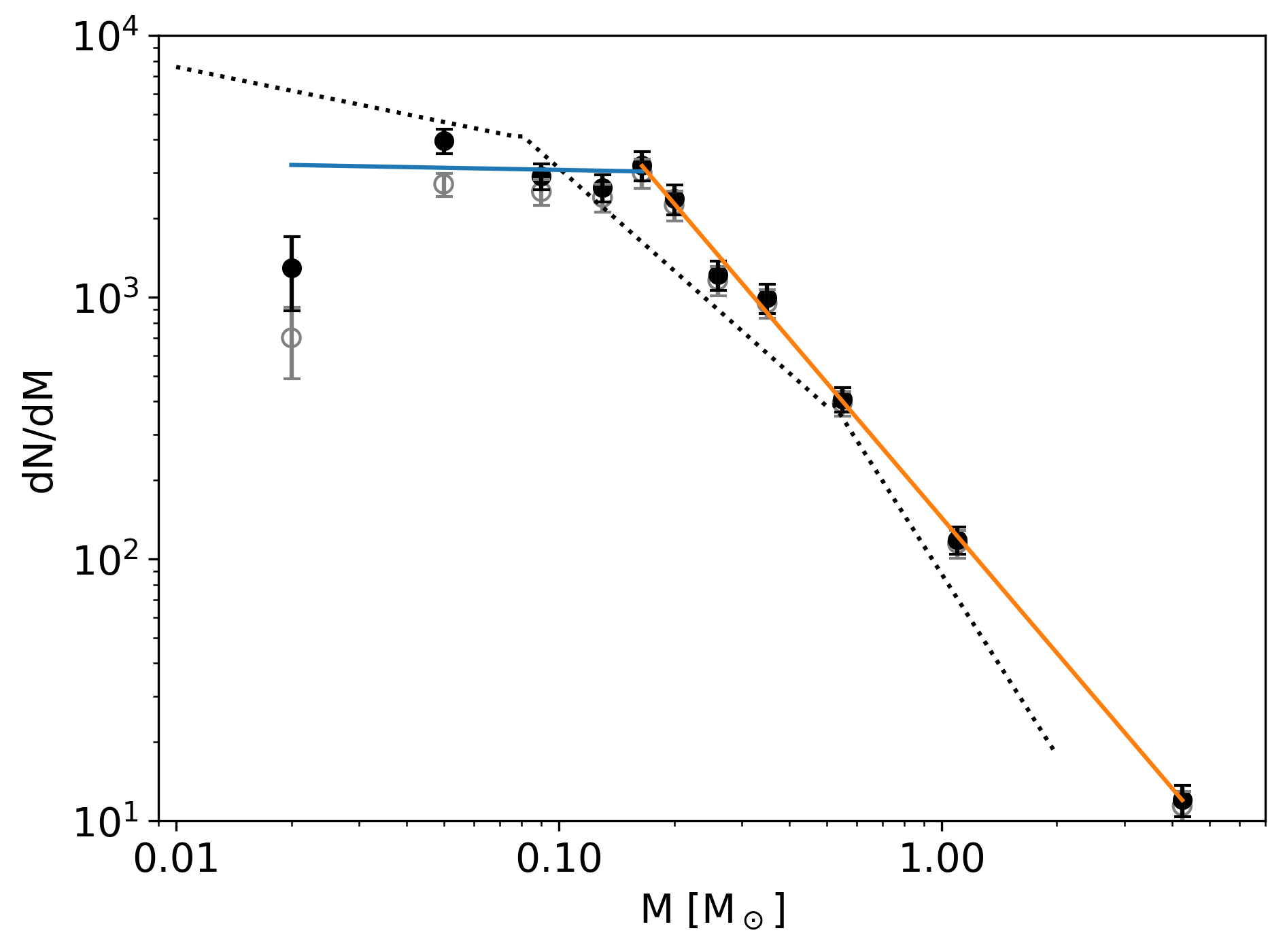}%{imf_model.png}
       % \caption{}
        \label{fig:imf-model}
    \end{subfigure}
%\hspace{-1}
    \begin{subfigure}{0.47\textwidth}
        \includegraphics[width=\textwidth]{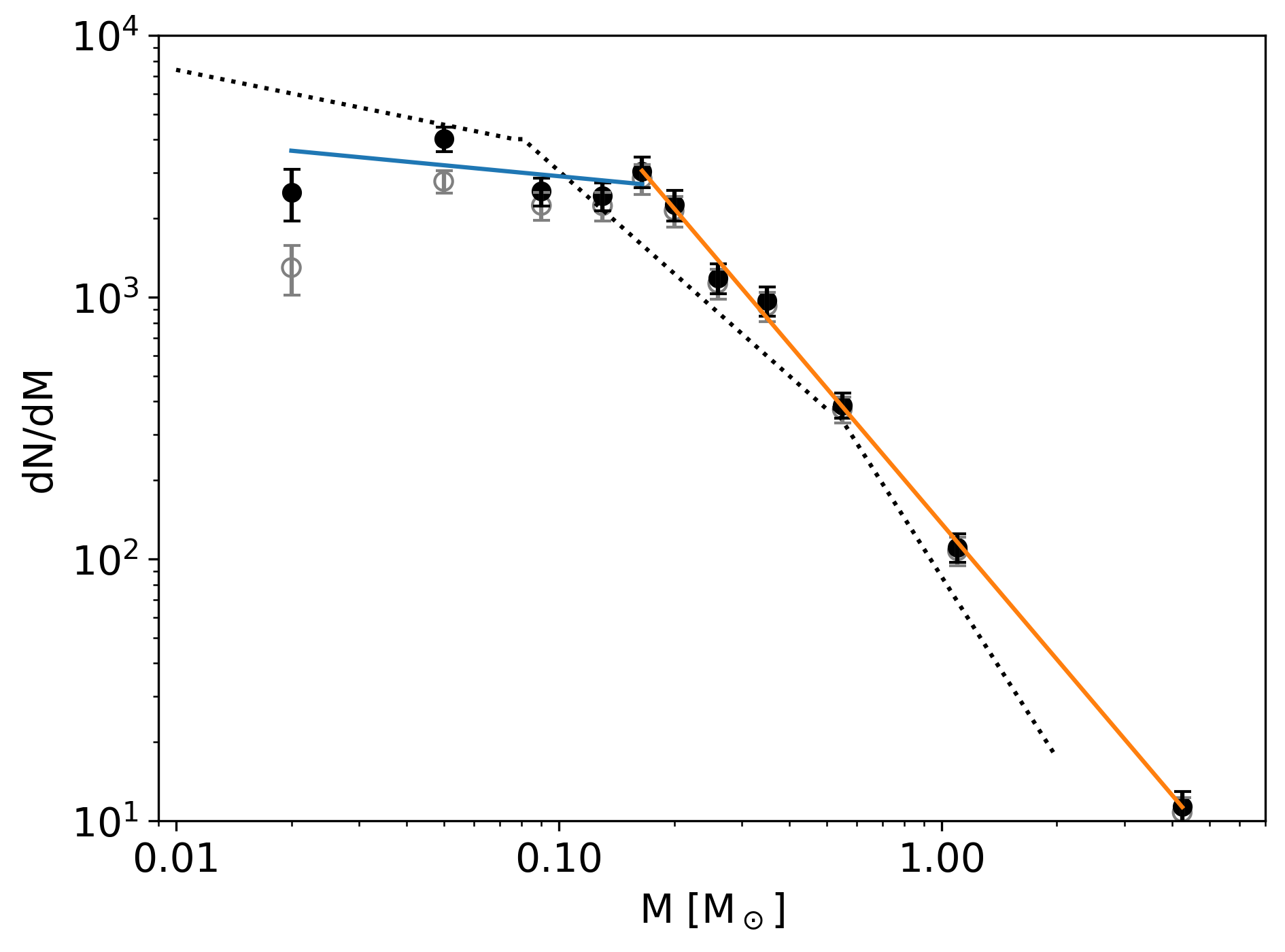}%{imf_vista.png}
        %\caption{}
        \label{fig:imf-vista}
    \end{subfigure}
\caption{d$N$/d$M$ representation of the IMF of Tr 14 after decontaminating with the Besançon model (left panel) and the VISTA (right panel) CF. Black dots with corresponding uncertainties are  after correcting for completeness, while grey circles are the bins before the correction. The black dotted line represents the standard mass function from \cite{kroupa2001}. The power-law break is at $\sim 0.2$ $M_\odot$. For the blue solid line fit in the left panel, we get the slope of  {$\alpha=0.03\pm0.24$}, and for the orange solid line  {$\alpha=1.72\pm0.04$}. For the blue solid line fit in the right panel we get the slope of  {$\alpha=0.14\pm0.19$} and for the orange solid line  {$\alpha=1.72\pm0.04$}.}
\label{fig:imf}
\end{figure*}

The results for the IMFs are shown in Fig. \ref{fig:imf} and summarized in Table \ref{tab:imf_results}. The left panel represents the IMF after decontaminating the cluster with the Besançon model, while the right panel after decontaminating with the VISTA CF. In both cases, we give the d$N$/d$M$ representation of the IMF of Tr 14. 
The black dotted line is the standard mass function from \cite{kroupa2001}. The black points with their uncertainties are the numbers of objects after the correction for completeness, while the grey circles with their uncertainties are the numbers of objects before the correction. 
The blue and orange solid lines are power-law fits with a break at 0.15 $M_\odot$. The obtained slope $\alpha$ ($dN/dM$ $\propto M^{-\alpha}$) when fitting between 0.01-0.2 $M_\odot$ differs whether we do the decontamination with the Besançon model or the VISTA CF: using the Besançon model, we get  {$\alpha = 0.03 \pm 0.24$}, whereas using the VISTA CF, we obtain  {$\alpha = 0.14 \pm 0.19$}. When fitting the IMF between $0.2-4.5\ M_\odot$ we find  {$\alpha = 1.72 \pm 0.04$} for both decontamination methods.

\begin{figure*}
\centering
    \begin{subfigure}{0.47\textwidth}
        \includegraphics[width=\textwidth]{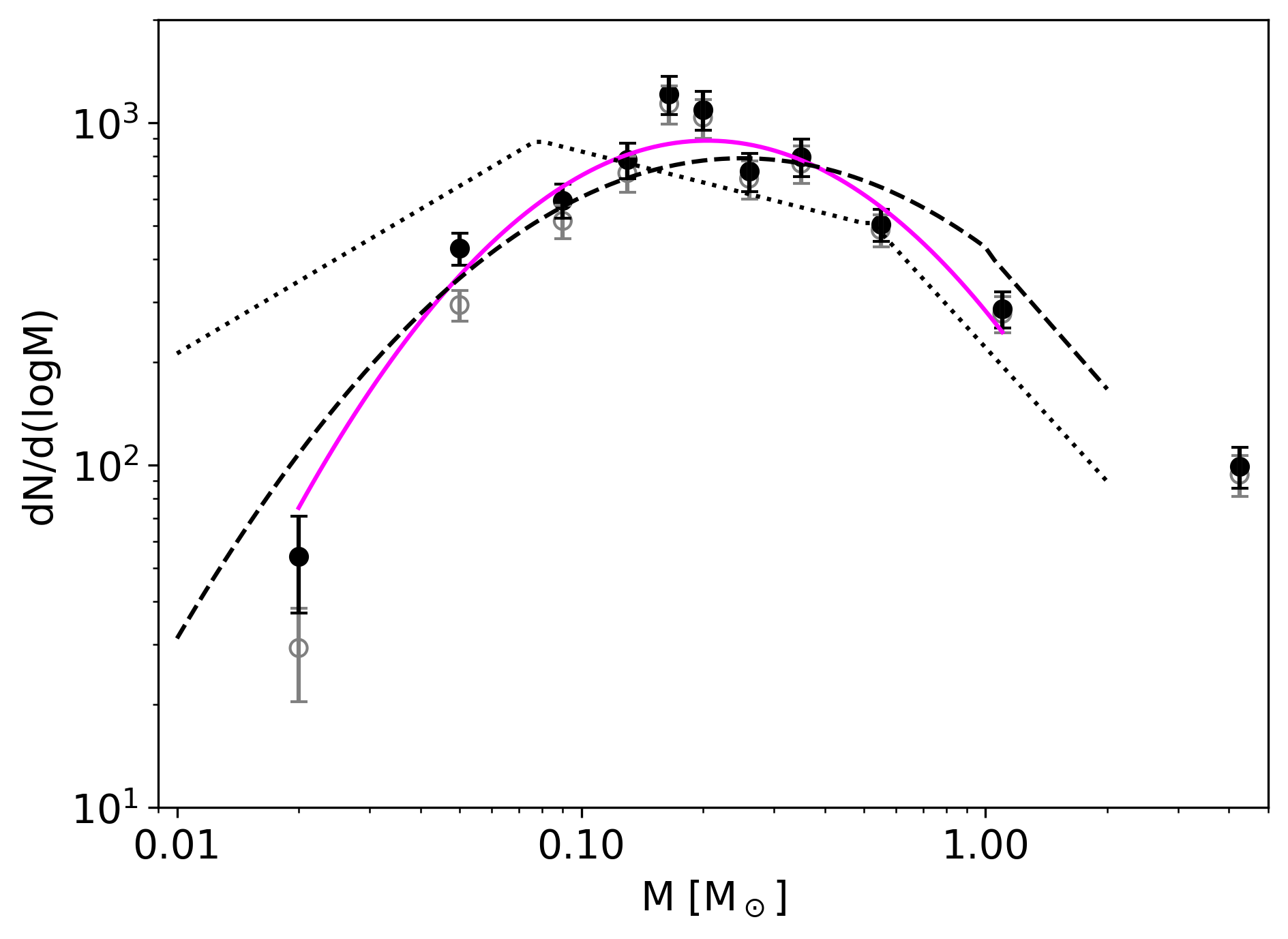}
       % \caption{}
        \label{fig:imf-model-log}
    \end{subfigure}
%\hspace{-1}
    \begin{subfigure}{0.47\textwidth}
        \includegraphics[width=\textwidth]{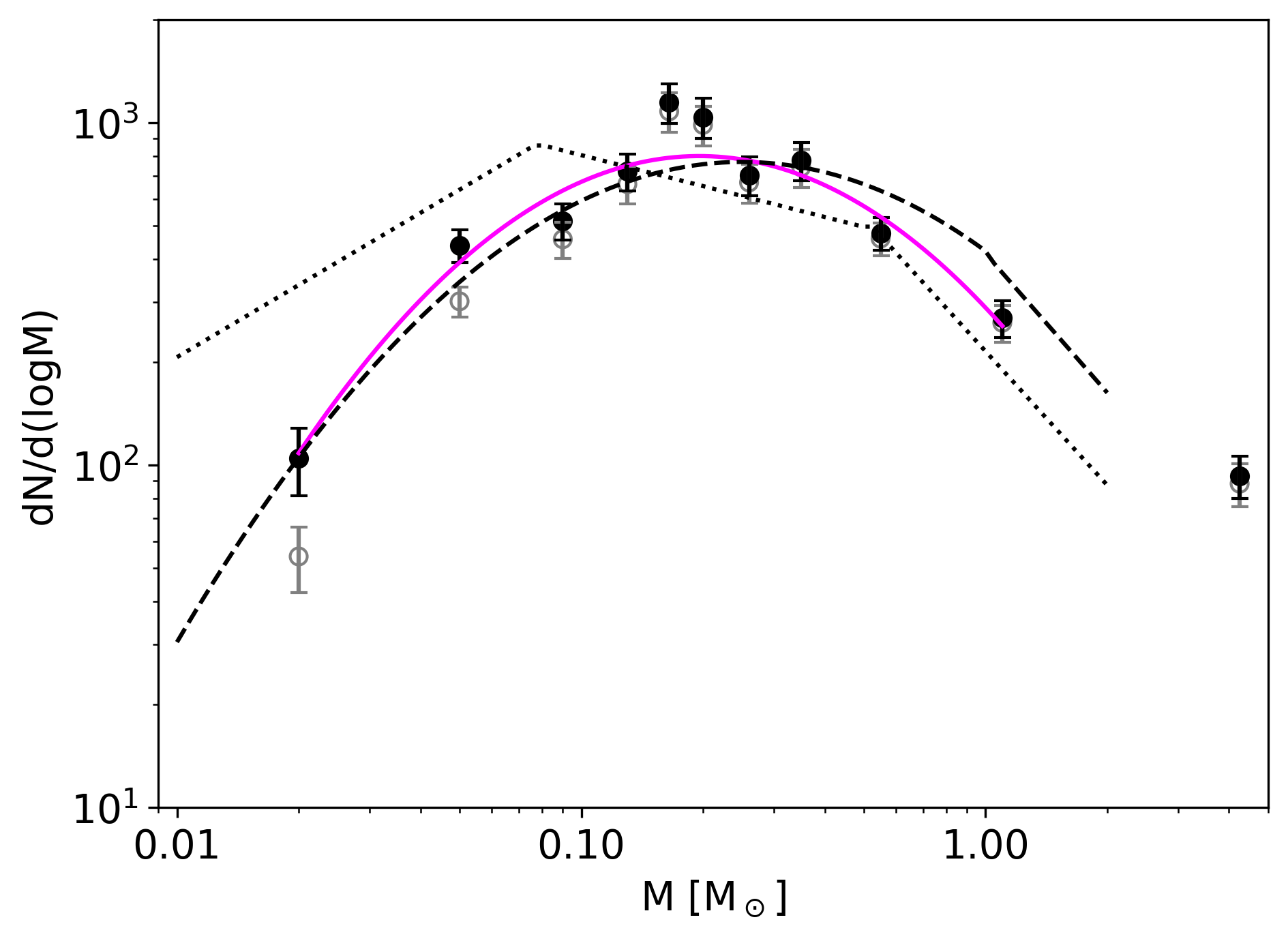}
        %\caption{}
        \label{fig:imf-vista-log}
    \end{subfigure}
\caption{d$N$/d(log$M$) representation of the IMF of Tr 14 after decontaminating with the Besançon model (left panel) and the VISTA (right panel) CF. Black dots with corresponding Poisson uncertainties are the result after correcting for completeness, while grey circles represent the IMF before the correction. The black dotted line represents the standard mass function from \cite{kroupa2001}, the black dashed line the mass function from \cite{chabrier2005} and the magenta solid line represents our log-normal fit of the IMF.}
\label{fig:imf-log}
\end{figure*}

 {We note that the decontamination primarily affects the lowest-mass bin, whereas above this limit the results remain very similar. On} the low-mass part of the IMF, the Besançon model may be 
overestimating the number of contaminants. This could originate from Tr 14 being a cluster located in a young star-forming region, implicating the presence of nebulosity and extinction. The approach with VISTA as the CF is more realistic because it directly samples the actual background and environmental conditions, whereas the Besançon model relies on assumptions that may not fully take into account local conditions.  {In particular, the model does not include explicit spiral-arm overdensities along this line of sight, such as the Carina arm, which could also cause it to underestimate the number of field stars near the cluster distance. However, even if such an overdensity were included, it would likely be uniformly distributed across the CMD and thus would not affect our decontamination result, especially in the lowest-mass bin where the largest difference relative to the VISTA CF is observed.} \cite{meingast2016} finds that the Besançon model overestimates number of faint (K$_S$ > 19 mag) and bright (K$_S$ < 10 mag) objects which could explain the results that we find. However, our approach with the VISTA CF could, due to its position, sample different parts of the Milky Way and thus potentially not see general Carina contamination, in addition to underestimating the foreground contamination, since Gaia does not have any parallax measurement for faint objects as explained in Sect. \ref{sec:vista}.  {Overall, the true number of contaminants, and consequently the resulting IMF slope, likely lies between the values inferred from the VISTA and Besançon model decontamination approaches, with both CFs yielding consistent results when the lowest-mass bin is excluded.}

In the dN/dlogM representation of the IMF \citep{chabrier2003}, as shown in Fig. \ref{fig:imf-log}, we fit a log-normal function. We derive a characteristic mass $m_c=(0.20\pm0.02)\ M_\odot$ and $\sigma=0.45\pm0.03$ for the decontamination approach using the Besançon model, while for VISTA we get $m_c=(0.20\pm0.02)\ M_\odot$ and $\sigma=0.50\pm0.03$. \cite{rochau} studied the central $\sim$1x1 arcmin$^2$ of Tr 14. They report a higher characteristic mass of $\sim0.52\ M_\odot$. For the dN/dM power law, they find $\alpha=0.37\pm0.32$ for masses $0.25-0.5\ M_\odot$ and $\alpha=1.52\pm0.3$ for masses $0.5-3.2\ M_\odot$. In contrast to \cite{rochau}, we do not find statistically significant evidence for a break in the IMF slope around 0.5 M$_\odot$. 
This may be due to our extended mass coverage and improved completeness, especially below 0.5 M$_\odot$.  Compared to \cite{damian2021}, who find $m_c=(0.32\pm0.02)\ M_\odot$ and $\sigma=0.47\pm0.02$ from a sample of nine clusters, our characteristic mass is lower, while the width of the distribution ($\sigma$) is similar or slightly broader. This shift toward lower $m_c$ may reflect differences in environmental conditions specific to Tr 14, or it may result from our deeper completeness and sensitivity to lower masses, particularly in the BD regime.

\begin{figure}
    \centering
\includegraphics[width=\linewidth]{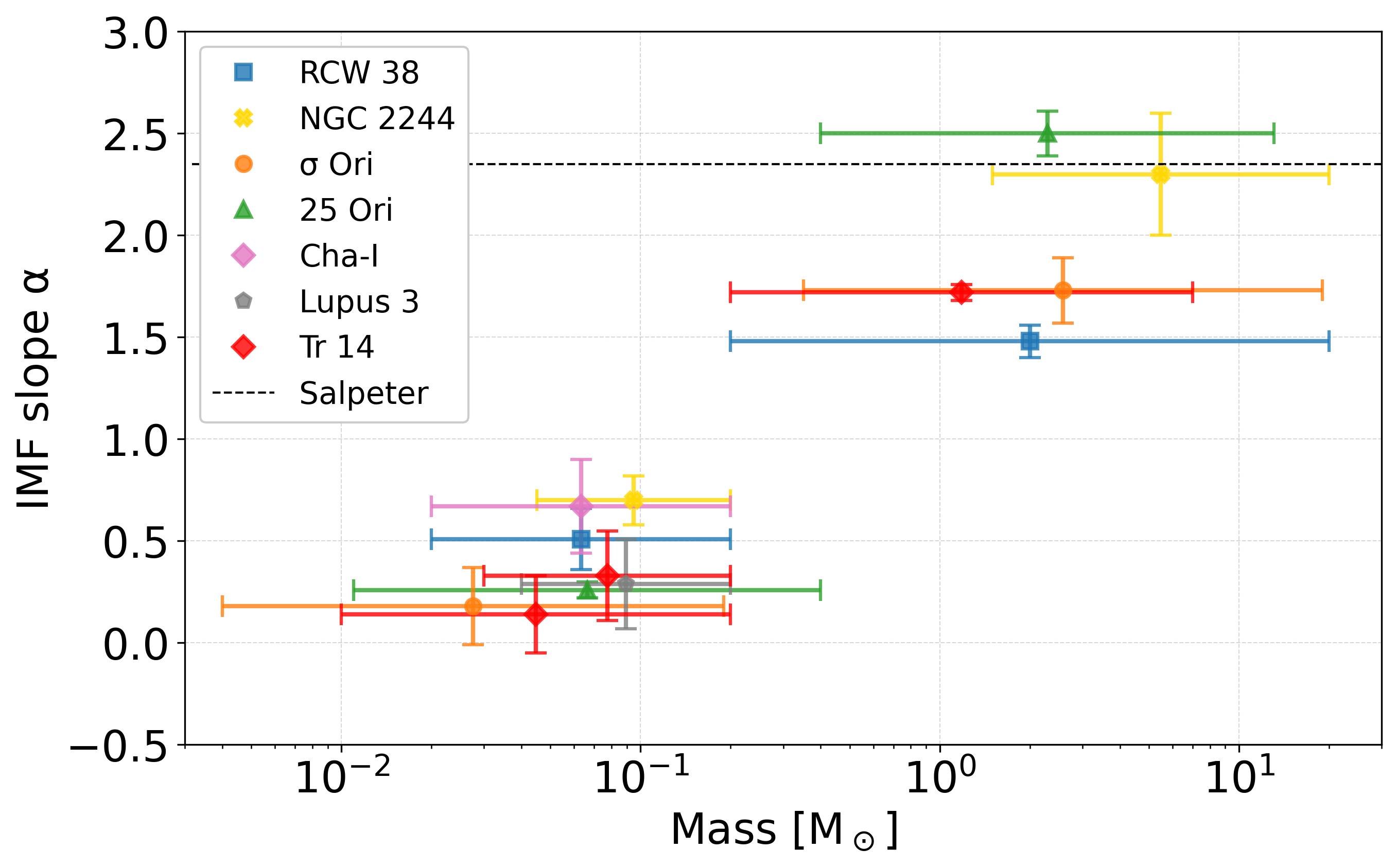}
    \caption{ {Alpha plot for the studies we compare our results to in the discussion.}  Each region is marked in different color and symbol as given in the legend, with its corresponding error bars for the $\alpha$. Tr 14 values are from this study. The two Tr 14 points at low masses correspond to the two different mass ranges used to compute the IMF slope. Horizontal bars represent the mass ranges over which the corresponding $\alpha$ values were derived and do not indicate uncertainties. }
    \label{fig:alpha}
\end{figure}

The results we obtain for the high mass part are similar to other studies in this mass range: for RCW 38, $\alpha=1.48\pm0.08$ for masses $0.2-20\ M_\odot$ \citep{muzic2017}; for $\sigma$ Ori, $\alpha=1.73\pm0.16$ for masses $0.35-19\ M_\odot$  \citep{penaramirez_2012}; but shallower than: 25 Ori where $\alpha=2.50\pm0.11$ for masses $0.4-13.1\ M_\odot$ \citep{suarez2019}, NGC 2244 where $\alpha=2.3\pm0.3$ for masses $1.5-20\ M_\odot$ \citep{muzic_stellar_2022} and for the standard Salpeter slope ($\alpha=2.35$). Our derived substellar IMF slope is significantly shallower than the typical values reported in the literature. For example, $\alpha=0.67\pm0.23$ in Cha-I for masses $0.02-0.2\ M_\odot$ and $\alpha=0.51\pm0.15$ in RCW 38 for masses $0.02-0.2\ M_\odot$  {\citep{corona}}; $\alpha=0.7\pm0.12$ for masses $0.045-0.2\ M_\odot$ in NGC 2244 \citep{victor2023}. Compared to these regions, our results suggest a flatter distribution of substellar objects, implying a lower relative BDs with respect to stars. However, we find  {$\alpha=0.33\pm0.22$} and  {$\alpha=0.26\pm0.16$} for the VISTA and Besançon model CF in the $0.03-0.2\ M_\odot$ mass range, similar to $\alpha=0.29\pm0.22$ in Lupus 3 for masses $0.04-0.2\ M_\odot$  {\citep{corona}}, $\alpha=0.26\pm0.04$ for masses $0.011-0.4\ M_\odot$ in 25 Ori \citep{suarez2019} and $\alpha=0.18\pm0.19$ for masses $0.004-0.19\ M_\odot$ in $\sigma$ Ori \citep{damian2023}. 
While our lowest-mass range $0.01-0.2\ M_\odot$ yields particularly shallow slopes, this result is based on the least massive point, where the incompleteness is highest and uncertainties are largest. If this point is excluded, the remaining slope values are consistent with those reported for other young clusters. 
The results are illustrated in Fig. \ref{fig:alpha}. See, e.g., \cite{hennebelle2024physicaloriginstellarinitial} for a more comprehensive summary. 

\begin{table}[]
    \caption{IMF slopes $\alpha$ (where $dN/dM \propto M^{-\alpha}$) derived for the combined BT-Settl and PARSEC evolutionary model, and different mass ranges, and for the two decontamination strategies.}
    \centering
    \begin{tabular}{cccc}
    \hline \hline
        CF & $0.01$--$0.2\ M_\odot$ & $0.03$--$0.2\ M_\odot$ & $0.2$--$4.5\ M_\odot$ \\
        \hline
        VISTA &  {$0.14 \pm 0.19$} &  $0.33 \pm 0.22$ &  $1.72 \pm 0.04$  \\
        Besançon model &  $0.03 \pm 0.24$ &  $0.26 \pm 0.16$ &  {$1.72 \pm 0.04$}  \\
        \hline
    \end{tabular}
    \label{tab:imf_results}
\end{table}

\begin{table*}[]
        \caption{IMF slopes $\alpha$ (where $dN/dM \propto M^{-\alpha}$) derived for different evolutionary model combinations in the $0.2$--$1.2\ M_\odot$ mass range.}
    \centering
    \begin{tabular}{cccc}
    \hline \hline
       CF & BT-Settl+PARSEC & PARSEC & BT-Settl \\
        \hline
        VISTA &  $1.70 \pm 0.20$ &  $1.58\pm0.21$ &  $1.53\pm0.15$  \\
        model &  $1.66\pm 0.19$ &  $1.55 \pm 0.20$ &  $1.49\pm0.13$ \\
        \hline
    \end{tabular}
    \label{tab:comparison}
\end{table*}

\subsection{Testing IMF assumptions}
\subsubsection{Dynamical evolution}
To interpret the observed mass function in our field of view as IMF, it is necessary to assume that no significant fraction of cluster members resides outside our observed region, and that dynamical evolution has not substantially altered the mass distribution since formation. While such assumptions are generally justified for very young clusters, Tr 14 is both massive and compact, with a half-mass radius of $1.5-1.8$ pc and a total stellar mass of several thousand solar masses \citep{shull2021gaia}. Its crossing time is estimated at $\sim0.7-0.8$ Myr, and its two-body relaxation time ranges from 40 to 80 Myr \citep{shull2021gaia}. Given the cluster’s age of $\sim1$ Myr, this places it just beyond one crossing time but still far from being dynamically relaxed. This suggests that dynamical effects such as the ejection of members or significant radial redistribution (especially for low-mass objects) are unlikely to have strongly influenced our sample. While \citet{sana2010} report signs of mass segregation, most likely due to dynamical friction, among stars more massive than $\sim$ 10 $M_\odot$, \citet{shull2021gaia} argue that even the most massive stars (30–80 $M_\odot$) are unlikely to have undergone significant segregation within the cluster’s 1–3 Myr lifetime. While this supports the idea that our sample, limited to lower-mass stars and BDs in the central region, is likely to still reflect the IMF, some degree of primordial mass segregation cannot be ruled out. A wider field study would be needed to verify whether spatially dependent effects like mass segregation influence the mass function at larger radii.

\subsubsection{Crowding}
To assess the potential influence of crowding and the halos of bright stars on the derived IMF we compared the IMFs extracted from two distinct regions within our field of view. We defined the most crowded region by drawing a rectangle over the central area of the image, while the rest of the field encompassing less crowded surroundings was treated as a separate region for comparison. We computed the photometric completeness separately for each region and subsequently derived their respective IMFs. We found only small differences between the two IMFs. For the mass range $0.01-0.2\ M_\odot$, we measured slopes of $ {\alpha_{\mathrm{VISTA},inside}=0.07\pm0.18}$ and similarly  {$\alpha_{\mathrm{VISTA},outside}=0.16\pm0.22$}, while for the Besançon model CF  {$\alpha_{\mathrm{model},inside}=-0.07\pm0.22$} and  {$\alpha_{\mathrm{model},outside}=0.05\pm0.26$}. In the mass range $0.2-4.5\ M_\odot$, we find  {$\alpha_{\mathrm{VISTA},inside}=1.68\pm0.06$} and  {$\alpha_{\mathrm{VISTA},outside}=1.76\pm0.04$}, while for the Besançon model CF  {$\alpha_{\mathrm{model},inside}=1.65\pm0.05$} and  {$\alpha_{\mathrm{model},outside}=1.77\pm0.05$}. These results indicate that crowding and the presence of bright stellar halos do not significantly bias our IMF determination.

\subsubsection{Evolutionary models}
To be able to cover the full range from the substellar to high stellar masses, in our analysis we used a BT-Settl and PARSEC isochrones combined into a single isochrone at 1 $M_\odot$. To estimate the impact of this combination, we evaluate the IMF with the two isochrones separately, within the common mass range. The results are given in Table \ref{tab:comparison}. We see that with the PARSEC \citep{bressan2012parsec} evolutionary model we have an overabundance of low-mass sources in comparison to the results obtained using BT-Settl \citep{baraffe2015new} evolutionary models. Using the PARSEC isochrone, in the derivation process of masses, large portion of objects 
will be crossing the lowest point on the isochrone, creating the artificial abundance of low-mass objects. We, thus, conclude that using a combined isochrone does not produce any anomalous effects at the intersection of the isochrones, and that the IMF is consistent with either of the two models.

\subsection{Star-to-BD ratio}
\label{sec:bd}
To calculate the star-to-BD ratio we use the same method as in the IMF derivation, up to the point where we bin the data. Instead, we count the number of sources in the mass ranges $0.03-0.075\ M_\odot$ for the BDs and $0.075-1\ M_\odot$ for the stars and correct for completeness, restricting only to sources with completeness >= 20\%. The median star-to-BD that we obtain using the VISTA CF is 4.0, with a 95\% confidence interval of 2.8-5.8, while for the Besançon model it is 4.3 with the 95\% confidence interval of 3.1-6.0. As stated previously, we consider the result with the VISTA CF to be more realistic. This result is in agreement with the star-to-BD ratios in RCW 38 ($2.0\pm0.6$); Cha-I ($3.2-4.8$) and Lupus 3 ($2.1-4.5$) from  {\citet{corona}}. Our result is in agreement with \cite{andersen_2008} which reports star-to-BD ratios in different regions in the range between $3.3-8.5$.

To place Tr 14 in context of different environments, we estimate its surface density and FUV flux from massive stars. We calculate a mean surface density of $\sim1000$ stars/pc$^{2}$ by dividing the number of sources in our field of view by the area. To estimate the FUV radiation, we do it identically as in  {\cite{corona}}. We derive a flux of approximately $10^{5.54}\ G_0$.

The Fig. 11 in \cite{victor2023} shows that the star-to-BD ratio does not significantly vary with stellar surface density, nor with the presence of OB stars. Their analysis includes clusters with surface densities in $\sim10-1000$ stars pc$^{-2}$ range, including ONC, RCW 38, NGC 2244, Lupus 3, Cha-I, NGC 1333 and IC 348. 
Tr 14 is on the high end of this range, placing it among the densest clusters examined and only surpassed by RCW 38.

In terms of FUV radation, Fig. 8 of  {\cite{corona}}, presents the median FUV flux for Lupus 3, Cha-I, NGC 2244, CrA, NGC 1333, and RCW 38.  Tr 14, with its estimated flux of $10^{5.54}\ G_0$, exceeds all the clusters in their sample, including RCW 38 ($\sim10^{5.3}\ G_0$).

Despite this extreme environment, high density and high FUV flux, our measured star-to-BD ratio evaluated for the mass range $0.03-0.075\ M_\odot$, lies between $\sim$3 and 6. This range is comparable to those observed in both low- and high-density environments in \cite{victor2023} and  {\cite{corona}}, suggesting that the star-to-BD ratio in Tr 14 does not significantly differ from other regions in this mass range. However, we note that the IMF we derive is significantly flatter than those in the comparison regions. This apparent contradiction can be explained when considering the slope uncertainties and excluding the lowest-mass bin, which drives much of the flattening; with that bin removed, the slope becomes more consistent with other clusters. However, if we were to include lower-mass BDs ($<0.03\ M_\odot$), where our IMF shows a clear deficit, we would likely start to see deviations from the trend, indicating a possible environmental influence on the formation efficiency of the lowest-mass substellar objects. Interestingly, while \citet{victor2023} propose that OB stars may enhance BD formation in NGC 2244, our results from the denser and more UV-intense environment of Tr 14 do not support this scenario, pointing instead to a potential suppression. It may be that a combination of high density and intense radiation fields inhibits the fragmentation processes needed for efficient formation of very low-mass objects. 

\section{Summary and conclusions}
\label{sec:fin}
Based on GSAOI/GeMS observations, we have presented the most complete mass function in Tr 14 to date, covering objects from the substellar regime up to 4.5 $M_\odot$.  Our final catalogue contains 1207 sources in $J$, $H$ and K$_S$ bands. The data is 50\% complete down to 0.02 $M_\odot$ and 20\% complete down to 0.01 $M_\odot$. Because we see both field stars and cluster members in our dataset, we statistically decontaminate the catalogue using either the Besançon model or observational data from the VISTA telescope as a CF. Masses and extinctions are obtained from the comparison to the evolutionary models from \cite{bressan2012parsec, chen2014improving, baraffe2015new}.The resulting IMF below 0.2 $M_\odot$ differs whether we use the Besançon  model or VISTA for the decontamination process. We believe that the Besançon model overestimates the contamination, and therefore prefer the results obtained using the observational CF.

Using the VISTA CF, we find the slope of the high-mass part of the IMF ($\sim0.2-4.5\ M_\odot$) to be  $\alpha=1.72\pm0.04$, which is similar to other young star-forming regions, but shallower than the standard Salpeter slope. For the low-mass end of the IMF ($\sim 0.01$–$0.2\ M_\odot$), we find a relatively shallow slope of  $\alpha = 0.14 \pm 0.19$. While this appears significantly flatter than values reported for other regions, the comparison should be treated with caution: not only do we probe a slightly different mass range than most studies, but the lowest-mass bin in this mass range is also the most affected by incompleteness, leading to larger uncertainties. If the least-massive point is excluded, the remaining slope values are consistent with those reported for other young clusters. 

Nonetheless, the observed flattening could suggest that the number of brown dwarfs declines steeply with decreasing mass, especially below 0.03 $M_\odot$, suggesting a lower formation efficiency for substellar objects in this environment. The cause of this suppression is unclear, but may relate to the specific environmental conditions of Tr 14. Its mean surface density of $\sim1000$ stars/pc$^{-2}$ lies between that of NGC 2244 (lower) and RCW 38 (higher), yet the substellar IMF we find is much flatter than in either of those regions. In terms of FUV flux, Tr 14 is exposed to stronger radiation fields than both NGC 2244 and RCW 38, which may inhibit the formation or survival of the lowest-mass BDs. Though we cannot yet conclude this definitively, our findings highlight the importance of considering both stellar density and radiative feedback in shaping the low-mass end of the IMF.

Finally, we show the star-to-BD ratio is 4.0 with the 95\% confidence interval of 2.8-5.8, in the mass range $0.03-1\ M_\odot$, which is within the expected range of $\sim$3-6 found in the literature. This places Tr 14 near the middle of the observed distribution of star-to-BD ratios, suggesting that in this specific mass range, BDs form at least as efficiently as low-mass stars, consistent with other environments. However, our IMF shows a significantly flatter slope and a deficit of objects at even lower masses (below 0.03 $M_\odot$), which are not included in this ratio. This may indicate that environmental effects primarily influence the formation efficiency of the lowest-mass BDs.
\begin{acknowledgements}
This work was supported by/supported in part by the Croatian Science Foundation under the project number HRZZ-MOBDOK-2023-9579.
K.M. acknowledges support from the Fundação para
a Ciência e a Tecnologia (FCT) through the CEEC-individual contract
2022.03809.CEECIND, and the Scientific Visitor Programme of the European
Southern Observatory (ESO) in Chile. V.A-A acknowledges support from the INAF grant 1.05.12.05.03.
Based on observations (program ID GS-2019A-DD-107, PI: M. Andersen) obtained at the international Gemini Observatory, a program of NSF NOIRLab, which is managed by the Association of Universities for Research in Astronomy (AURA) under a cooperative agreement with the U.S. National Science Foundation on behalf of the Gemini Observatory partnership: the U.S. National Science Foundation (United States), National Research Council (Canada), Agencia Nacional de Investigaci\'{o}n y Desarrollo (Chile), Ministerio de Ciencia, Tecnolog\'{i}a e Innovaci\'{o}n (Argentina), Minist\'{e}rio da Ci\^{e}ncia, Tecnologia, Inova\c{c}\~{o}es e Comunica\c{c}\~{o}es (Brazil), and Korea Astronomy and Space Science Institute (Republic of Korea).

This work has made use of data from the European Space Agency (ESA) mission
{\it Gaia} (\url{https://www.cosmos.esa.int/gaia}), processed by the {\it Gaia}
Data Processing and Analysis Consortium (DPAC,
\url{https://www.cosmos.esa.int/web/gaia/dpac/consortium}). Funding for the DPAC
has been provided by national institutions, in particular the institutions
participating in the {\it Gaia} Multilateral Agreement.

\end{acknowledgements}

\bibliographystyle{aa} 
\bibliography{aa56578-25}

\begin{appendix}
\section{Natural guide stars}
\label{appendix:ngs}

In Table~\ref{tab:ngs}, we list the positions of the natural guide stars.

\begin{table}[h]
    \caption{Labels and positions of the natural guide stars.}
    \centering
    \begin{tabular}{ccc}
    \hline \hline
    GWFS guide-star ID & RA & Dec\\
    \hline
    '153-055026' & 10:44:02.72 & –59:32:28.36
    \\
    '153-054974' & 10:43:57.57 & –59:33:38.66 
    \\
    '153-054913' & 10:43:52.06 & –59:32:40.28 
    \\
    \hline
    \end{tabular}
    \label{tab:ngs}
\end{table}

\section{Photometry parameters}
\label{appendix:parameters}

In Table~\ref{tab:parameters}, we list the parameters used in the extraction of photometry using {\sc{SourceExtractor}} $\mathrm{and}$ {\sc{PSFEx}}.

\begin{table}[b]
\centering
\caption{Parameters used for the photometric extraction. \label{tab:parameters}}
\begin{tabular}{lcccccc} \hline \hline \\
& \multicolumn{6}{c}{long exposure}
\\
 & \multicolumn{3}{c}{extraction 1} &\multicolumn{3}{c}{extraction 2} \\
 parameter & $J$   & $H$ & ${\mathrm{K}_S}$ &  $J$   & $H$ & ${\mathrm{K}_S}$ \\
\hline
BACK\_SIZE & 64 & 64 & 64 & 128 & 128 & 128 \\
BACK\_FILTERSIZE & 3  & 3 & 3 & 5 & 5 & 3\\
DETECT\_THRESH &7 &7 & 7 & 1.5&1.5&1.5\\
\hline
BASIS\_NUMBER  & \multicolumn{6}{c}{20} \\
PSF\_SIZE & \multicolumn{6}{c}{35,35} \\
PSFVAR\_KEYS & \multicolumn{6}{c}{X\_IMAGE, Y\_IMAGE} \\
PSFVAR\_GROUPS & \multicolumn{6}{c}{1,1} \\
PSFVAR\_DEGREES & \multicolumn{6}{c}{3} \\
SAMPLE\_MINSN & \multicolumn{6}{c}{20} \\
\hline
SPREAD\_MODEL &$0.003\pm 0.167$ &$0.004\pm 0.017$ &$0.003\pm 0.016$ &$0.002\pm0.024$ &$0.003\pm0.025$&$0.003\pm 0.027$\\
\hline
\hline
\\
& \multicolumn{6}{c}{short exposure}
\\
 & \multicolumn{3}{c}{extraction 1} &\multicolumn{3}{c}{extraction 2} \\
 parameter & $J$   & $H$ & ${\mathrm{K}_S}$ &  $J$   & $H$ & $\mathrm{K}_S$ \\
\hline
BACK\_SIZE & 64 & 64 & 64 & 128 & 128 & 128 \\
BACK\_FILTERSIZE & 3  & 3 & 3 & 3 & 5 & 5\\
DETECT\_THRESH &7 &7 & 7 & 1.5&1.5&1.5\\
\hline
BASIS\_NUMBER  & \multicolumn{6}{c}{20} \\
PSF\_SIZE & \multicolumn{6}{c}{35,35} \\
PSFVAR\_KEYS & \multicolumn{6}{c}{X\_IMAGE, Y\_IMAGE} \\
PSFVAR\_GROUPS & \multicolumn{6}{c}{1,1} \\
PSFVAR\_DEGREES & \multicolumn{6}{c}{3} \\
SAMPLE\_MINSN & \multicolumn{6}{c}{20} \\
\hline
SPREAD\_MODEL &$0.004\pm 0.020$ &$0.002\pm 0.011$ &$0.0002\pm 0.0091$ &$0.003\pm0.022$ &$0.001\pm0.014$&$0.0002\pm 0.0124$\\
 
\hline
\end{tabular}
\end{table}

\section{Besançon Galactic model photometric errors}
\label{appendix:besancon}

The expressions for photometric errors, in corresponding bands, given to the Besançon Galactic model for the expected foreground population:

\begin{align*}
&\sigma_J = 0.116 + \exp(1.15 \cdot J - 28.5) \\
&\sigma_H = 0.13 + \exp(1.3 \cdot H - 31) \\
&\sigma_{K_S} = 0.13 + \exp(1.3 \cdot K_S - 30).
\end{align*}

The photometric errors, for corresponding bands, for the background population are given:

\begin{align*}
& \sigma_J = 0.116 + \exp(c_J \cdot J_r - b_J) \\
& \sigma_H = 0.13 + \exp(c_H \cdot H_r - b_H) \\
& \sigma_{K_S} = 0.13 + \exp(c_{K_S} \cdot K{_S}_r - b_{K_S})
\end{align*}

\begin{table}[h!]
\centering
\caption{Coefficients $c$ and offsets $b$ for different values of $A_V$ \label{tab:coef} in the input of the Galactic model.}
\begin{tabular}{ccccccc}
\hline \hline
$A_V$ & $c_J$ & $c_H$ & $c_{K_S}$ & $b_J$ & $b_H$ & $b_{K_S}$ \\
\hline
5 & 1.20 & 1.29 & 1.32 & 28.0 & 30.0 & 30.0 \\
6 & 1.23 & 1.30 & 1.33 & 28.0 & 30.0 & 30.0 \\
7 & 1.23 & 1.31 & 1.34 & 27.5 & 30.0 & 30.0 \\
\hline
\end{tabular}
\end{table}

where the coefficients are given in Table \ref{tab:coef} for the specific extinction value that we use and the index $r$ notes that the values of the magnitudes used are reddened for the $A_V$.
\end{appendix}

\end{document}